\documentclass[11pt]{article}

\usepackage{amsfonts}
\usepackage{amsmath}
\usepackage{amssymb} 
\usepackage{amsthm}
\usepackage{authblk} 
\usepackage[a4paper, margin = 1.0in]{geometry} 
\usepackage{bm}
\usepackage{float}
\usepackage{graphicx} 
\usepackage[pagebackref=true, colorlinks=true, linkcolor=blue, citecolor=blue]{hyperref}
\usepackage{lmodern}
\usepackage{mathtools}
\usepackage[round]{natbib}
\usepackage{pifont} 
\usepackage[section]{placeins}
\usepackage{setspace}
\usepackage{tabularx,booktabs}
\usepackage{threeparttable}
\usepackage{url}
\usepackage{xcolor}
\usepackage{xr-hyper}

\floatstyle{plaintop}
\restylefloat{table}

\definecolor{cb-green-sea}  {RGB}{  0,146,146}
\definecolor{cb-burgundy}   {RGB}{146,  0,  0}
\definecolor{cb-green-lime} {RGB}{ 36,255, 36}
\newcommand{\cmark}{\color{cb-green-sea}\ding{51}}%
\newcommand{\xmark}{\color{cb-burgundy}\ding{55}}%


\renewenvironment{abstract}
 {\par\noindent\textbf{\abstractname}\ \ignorespaces}
 {\par\medskip}

\newtheorem{proposition}{Proposition}
\newtheorem{corollary}{Corollary}
\newtheorem{lemma}{Lemma}
\newcommand{\df}{f^{(1)}}
\newcommand{\dy}{y^{(1)}}
\newcommand{\bx}{\mathbf{x}}
\newcommand{\by}{\mathbf{y}}
\newcommand{\bdy}{\mathbf{y}^{(1)}}
\newcommand{\Bf}{\mathbf{f}}
\newcommand{\bg}{\mathbf{g}}
\newcommand{\bdf}{\mathbf{f}^{(1)}}
\newcommand{\dsigma}{\sigma^{(1)}}
\newcommand{\dalpha}{\alpha^{(1)}}
\newcommand{\brho}{\bm{\rho}}
\newcommand{\bsigma}{\bm{\sigma}}
\newcommand{\balpha}{\bm{\alpha}}
\newcommand{\bdsigma}{\bm{\sigma}^{(1)}}
\newcommand{\bdalpha}{\bm{\alpha}^{(1)}}

\title{\huge\textbf{Latent variable estimation with composite Hilbert space Gaussian processes}}

\author[1,2]{Soham Mukherjee \thanks{\texttt{soham.mukherjee@tu-dortmund.de}}}
\author[1]{Javier Enrique Aguilar}
\author[2,3]{Marcello Zago}
\author[2,3]{Manfred Claassen}
\author[1]{Paul-Christian B\"urkner}

\affil[1]{Department of Statistics, TU Dortmund University, Dortmund, Germany}
\affil[2]{Department of Computer Science, University of T\"ubingen, T\"ubingen, Germany}
\affil[3]{University Hospital T\"ubingen, Faculty of Medicine, University of T\"ubingen, T\"ubingen, Germany}
\date{} 

\begin{document}

\maketitle


\bigskip
\begin{abstract}
    We develop a scalable class of models for latent variable estimation using composite Gaussian processes, with a focus on derivative Gaussian processes. We jointly model multiple data sources as outputs to improve the accuracy of latent variable inference under a single probabilistic framework. Similarly specified exact Gaussian processes scale poorly with large datasets. To overcome this, we extend the recently developed Hilbert space approximation methods for Gaussian processes to obtain a reduced-rank representation of the composite covariance function through its spectral decomposition. Specifically, we derive and analyze the spectral decomposition of derivative covariance functions and further study their properties theoretically. Using these spectral decompositions, our methods easily scale up to data scenarios involving thousands of samples. We validate our methods in terms of latent variable estimation accuracy, uncertainty calibration, and inference speed across diverse simulation scenarios. Finally, using a real world case study from single-cell biology, we demonstrate the potential of our models in estimating latent cellular ordering given gene expression levels, thus enhancing our understanding of the underlying biological process.
\end{abstract}

\noindent%
\vfill

\section{Introduction} \label{sec-intro}
Gaussian processes (GPs) are a powerful class of non-parametric methods, allowing to tackle a wide range of research problems \citep{Rasmussen1995}. 
Latent variable estimation, a major branch in statistical modeling, is among these problems solvable with GPs and its extensions \citep{gp_rasmussen_williams_2006}. 
Latent variable GPs have not only been used for classification and dimension reduction \citep{lawrence_gaussian_2003, lawrence_probabilistic_2005, alvarez_latent_2009}, but also for regression models \citep{bodin2017latentgaussianprocessregression}. 
Other extensions of GPs include multi-output GPs \citep{teh_semiparametric_2005, bonilla_multi-task_2007} and derivative GPs \citep{solak_derivative_2002}.
Multi-output GPs, as the name suggests, use multi-dimensional (rather than uni-dimensional) data as their joint output. 
Derivative GPs, on the other hand, simultaneously model derivative information in addition to the original data through a joint derivative covariance function structure. 
Derivative GPs are a special case of jointly modeling two GPs with shared inputs. 
We will refer to this general framework as \textit{composite GPs}. 
While the naming of this general structure as composite GPs aren't popular (to the best of our knowledge), early uses of composite GPs can be found in, for example, specifying the joint predictive distribution of GPs \citep{gp_rasmussen_williams_2006}.

A commonality between both the multi-output and derivative GP extensions is to simultaneously model various sources of data in a single GP framework. 
Leveraging this aspect, \cite{mukherjee_dgp-lvm_2025} recently combined all of the aforementioned extensions for the purpose of latent variable estimation with multi-output derivative GPs, leading to drastic increase in accuracy for latent variable estimates. 
However, a major drawback of this framework lies in its applicability to large data, since exact GPs scale cubically with the number of observations. 
In this paper, we propose a scalable version of the composite GPs for latent variable estimation using multi-dimensional outputs that overcomes the scalability issues through the use of approximate GPs. 
As a special case, we will also show scalable  derivative GPs designed for the same purpose.

Various types of approximation methods for GPs have been discussed over the years. 
A major direction among them is to consider a reduced-rank representation of the gram matrix based on the chosen covariance function \citep{smola_sparse_2000, seeger_fast_2003, snelson_sparse_2005, quinonero-candela_unifying_2005}. 
In these works, the reduced rank representation is achieved by selecting a small set of representative samples known as inducing points. 
By solving the covariance function only on the inducing points, computation becomes much faster compared to inverting the gram matrix of the full sample. 

Another line of research under the reduced-rank representation approach focuses on Hilbert space approximate GPs (HSGPs) to speed up GP computation \citep{solin_hilbert_2020, riutort-mayol_approx_gps_2022}. 
Under the HSGP framework, the covariance function is approximated through its spectral decomposition which is computed from a finite set of representative basis functions. 
Exploiting the spectral representation of a stationary covariance function, HSGPs scale linearly with both sample size and the number of basis functions. 
Recently, HSGPs were extended to multi-output latent variable GPs in \cite{mukherjee2025hilbert} where the authors demonstrated their advantages in fast GP computation and estimating well-calibrated latent variables. 
Here, we further generalize the framework to composite GPs, with a specific focus on derivative GPs.

As an application of our developed framework, we tackle a contemporary research problem in single-cell biology where we estimate latent cellular ordering from single-cell RNA (scRNA) gene expression data \citep{haque_practical_2017}. 
This latent cellular ordering provides a proxy for the properties of the underlying biological processes \citep{trapnell_dynamics_2014} and is thus a crucial component that is to be estimated. 
To achieve this, we model unspliced and spliced gene expression data using composite GPs. 
Recently, latent cellular ordering were estimated based on RNA velocity \citep{la_manno_rna_2018, bergen_generalizing_2020}, which is obtained by estimating the rate of change between the spliced and unspliced gene expressions. Thus, we provide an alternative approach to estimating latent cellular ordering using spliced gene expression and RNA velocity through a scalable derivative GP model. ScRNA gene expression data is known for their large number of samples (cells) along with several inter-correlated genes.  
Concretely, we show how our proposed framework overcomes the aforementioned scalability issues and models full-sized scRNA gene expression data with thousands of cells, thus providing a practical approach of accurately estimating latent cellular ordering.

\subsection{Overview of contributions}
We develop a scalable class of composite GPs and its special case, derivative GPs for latent variable estimation. 
Based on our theoretical results, we provide a way to easily obtain spectral densities for a general derivative covariance function. 
We utilize these spectral densities to obtain an Hilbert space approximation for derivative GPs, as well as composite GPs more generally. 
Through extensive simulations, we show that our methods achieve accurate latent variable estimation scaling up to large data scenarios that are impractical for exact GPs. 
We demonstrate the potential of our methods in estimating latent cellular ordering from full-sized single-cell gene expression data, thus enhancing our understanding of the underlying biological process.

\section{Related work} \label{sec-rel-works}
In the field of latent variable modeling, GPs constitute a broad class of models. 
Latent variable GPs were introduced in \cite{lawrence_gaussian_2003, lawrence_probabilistic_2005} tasked towards dimensionality reduction via point estimation methods. 
Later works introduced estimating latent variable inputs using Bayesian inference \citep{titsias_bayesian_2010} as well as approximate variational inference methods for applications to high-dimensional data scenario \citep{hensman_gaussian_2013}.
Recently, a generalized latent variable estimation framework supporting various inference strategies were presented in \cite{lalchand_generalised_2022}. Other GP extensions include modeling multi-dimensional outputs \citep{teh_semiparametric_2005, bonilla_multi-task_2007} and simultaneous modeling of derivative information alongside GP outputs \citep{solak_derivative_2002}. Recently, all of the three above extensions were combined in a single modeling framework \citep{mukherjee_dgp-lvm_2025}, for which we propose a scalable version, among other contributions made in this paper.

While GPs are lauded for their flexible modeling capabilities, practical applications remain prohibitive for large sample sizes due to the computational requirements for exact GPs.  
Strategies like inducing points approximations \citep{smola_sparse_2000, snelson_sparse_2005, quinonero-candela_unifying_2005} combined with fast variational inference \citep{titsias_variational_2009, VIGPsWilk2016} have been developed to address the practical limitations of GPs. Recently, Hilbert space GPs (HSGPs) \citep{solin_hilbert_2020, riutort-mayol_approx_gps_2022} were developed where the covariance function is approximated through its spectral decomposition for a scalable solution. Furthermore, the multi-output latent variable HSGPs  \citep{mukherjee2025hilbert} promise well-calibrated, fast inference for latent variable estimates from large data scenarios. Here, we further generalize the HSGP framework to composite GPs, with a specific focus on derivative GPs \citep{mukherjee_dgp-lvm_2025}.

GPs have been popularly used in the field of single-cell biology \citep{buettner_novel_2012, hensman_hierarchical_2013, buettner_computational_2015}. Specifically, various works estimate the latent cellular ordering \citep{pseudotime_estimation_reid_john_wernisch, campbell_descriptive_2018} along with branching structures for trajectory inference \citep{ahmed_grandprix_2019} using GPs for understanding the true biological process. A limitation in these works lies in their restricted use of analyzing gene expression levels from a single source of information. In this paper, we showcase how incorporating additional information on gene expression levels results in an increased accuracy for these latent ordering estimates through our developed models.

\section{Methods} \label{sec-methods}
We generalize the HSGP framework to composite and derivative GPs for latent variable estimation. 
We first discuss the general composite GP structure jointly specified for two GPs along with the full derivative GPs as a special case. 
Following that, we propose the mathematical conditions for the composite (and derivative) GPs and develop a scalable method through Hilbert space approximations. 
As our setup, we will start by considering a pair of uni-dimensional output (response) variables $\by_f$ and $\by_g$ and a single (shared) input variable $\bx$. 
Accordingly, we will consider observations $(y_{f_i}, y_{g_i}, x_i)$ for $i = {1, 2, ..., N}$ where $N$ is the sample size. 
We will later extend the notations to consider multi-dimensional outputs and latent variable inputs.

\subsection{Composite Gaussian processes} \label{sec-comp-gp}
Consider scenarios where we want to simultaneously study two different sources of information through two stochastic processes $\Bf$ and $\bg$. 
We assume that these two processes individually follow a GP such that $f(x) \sim \mathcal{GP}(m_f, k_f)$ and $g(x) \sim \mathcal{GP}(m_g, k_g)$. 
The GPs are defined by mean functions $m_f = m_f(x)$, $m_g = m_g(x)$ and covariance functions $k_f = k_f(x,x')$ and $k_g = k_g(x, x')$, respectively, over the inputs $x, x' \in \mathbb{R}$. Jointly, we call them composite GPs. We specify them similar to the joint predictive GPs \citep{gp_rasmussen_williams_2006} and write the composite GP as
\begin{equation} \label{eqn-comp-gp}
    \left(\begin{matrix}
        f(x) \\
        g(x)
    \end{matrix}\right)
    \sim
    \mathcal{GP}\left(
    \left(\begin{matrix}
        m_f \\
        m_g
    \end{matrix}\right),
    \left(\begin{matrix}
        k_f & k_{fg}\\
        k_{gf} & k_g
    \end{matrix}\right)\right),
\end{equation}
where $k_{fg}$ and $k_{gf}$ are functions encoding the relationship (i.e., interactions) between $f(x)$ and $g(x)$, such that, $\text{Cov}(f_i, g_j) = k_{fg}(x_i, x_j)$ and $\text{Cov}(g_i, f_j) = k_{gf}(x_i, x_j)$.
In real world scenarios, the functional form of $k_{fg}$ and $k_{gf}$ is often unclear. 
In such cases, it becomes challenging to specify the full joint covariance structure of a composite GP model. The Hilbert space framework requires knowing the explicit functional forms of $k_{fg}$ and $k_{gf}$. 

The composite GPs have two (univariate) output variables $\by_f$ and $\by_g$. 
Modeling the common relationship of $\bx$ with $\by_f$ and $\by_g$ given the corresponding GPs with independent additive noises can be written as
\begin{equation} \label{eqn-comp-gp-model1}
    y_{f_i}  = f(x_i) + \varepsilon_{f_i} \quad \text{and} \quad  y_{g_i}  = g(x_i) + \varepsilon_{g_i},
\end{equation}
where $\varepsilon_{f_i} \sim \mathcal{N}(0, \sigma_f^2)$ and $\varepsilon_{g_i} \sim \mathcal{N}(0, \sigma_g^2)$. Together, this is equivalent to 
\begin{equation} \label{eqn-comp-gp-model2}
    y_{f_i} \mid f \sim \mathcal{N}(f(x_i), \sigma_f^2) \quad \text{and} \quad y_{g_i} \mid g \sim \mathcal{N}(g(x_i), \sigma_g^2).
\end{equation}
Thus, for $\by_f$, when $i \neq j$ we have $\text{Cov}(y_{f_i}, y_{f_j}) = k_f(x_i, x_j)$ and for $i=j$, we have $\text{Cov}(y_{f_i}, y_{f_j}) = \text{Var}(y_{f_i}) = k_f(x_i, x_j) + \sigma_f^2$. 
This is similar for $\by_g$ involving $k_g$ and $\sigma_g$ instead. 

Under this framework, the functional form of $k_f$ and $k_g$ needs to be chosen.
Although we show our theoretical results for a general covariance function, in this paper, we refer to the Mat\'{e}rn family and specifically the Squared exponential (SE) covariance functions \citep{gp_rasmussen_williams_2006}.
For practical purposes, we prefer SE since it is infinitely differentiable: a property which we use in our derivative GPs discussed below in Section \ref{sec-deriv-gp}.

\subsection{Derivative Gaussian processes} \label{sec-deriv-gp}
Using the framework of composite GPs, joint derivative GPs are a special case where we consider both $\Bf$ and its derivative $\bdf$ \citep{solak_derivative_2002, mukherjee_dgp-lvm_2025}.
The joint derivative GP is used when we wish to model data $\by$ and its derivative $\bdy$.
Concretely, we consider $f(x)$ and its derivative process $\df(x)$ such that $\df(x) \sim \mathcal{GP}(m_{\df}, k^{(1,1)})$ \citep{solak_derivative_2002, gp_rasmussen_williams_2006}, where $ k^{(1,1)}(x, x') = \partial_x^{1}\partial_{x'}^{1}k(x,x')$ denotes the second order partial derivative of $k(x, x')$ differentiated with respect to both $x$ and $x'$. We model $f(x)$ and $\df(x)$ jointly as 
\begin{equation}\label{eqn-deriv-gp}
    \left(\begin{matrix}
        f(x) \\
        \df(x)
    \end{matrix}\right)
    \sim
    \mathcal{GP}\left(
    \left(\begin{matrix}
        m_f \\
        m_{\df}
    \end{matrix}\right),
    \left(\begin{matrix}
        k & k^{(1,0)}\\
        k^{(0,1)} & k^{(1,1)}
    \end{matrix}\right)\right),
\end{equation}
where $k^{(1,0)} = \partial_x^{1}k(x,x')$ and $k^{(0,1)} =\partial_{x'}^{1}k(x,x')$. 
Thus, the off-diagonal covariances are obtained through the gram matrix $\mathbf{K}^{(1,0)}$ generated by function $k^{(1,0)}$ and its transpose $\mathbf{K}^{(0,1)}$ generated by $k^{(0,1)}$. 
Based on the aforementioned GP framework, we then model outputs $\by$ and its derivative $\bdy$ which can be obtained by replacing $\by_f$ and $\by_g$ in Eq. \ref{eqn-comp-gp-model2} by $\by$ and $\bdy$ respectively. 
The independent additive noises with error SDs $\sigma_f$ and $\sigma_g$ then correspond to $\by$ and $\bdy$ respectively.

Although we only show the second order partial derivatives in Eq. \ref{eqn-deriv-gp}, this structure holds for higher order derivative functions as well. Considering the general derivative structure of covariance functions $k$ and their corresponding spectral densities $S_k$ \citep{gihman_linear_2004} (see Section \ref{sec-hs-approx} for details) we propose the following result:

\begin{proposition}
\label{prop:derivative-kernel-main-text}
Let $k(x, x')= k(r)$, where $r = x - x'$, be a stationary covariance kernel on $\mathbb{R}$ 
with spectral density $S_k \in L^1 \left( \mathbb{R} \right)$ such that, for some integer $m\ge 1$,
\[
\int_{\mathbb{R}} \|\omega\|^m S_k(\omega)\,d\omega<\infty.
\]
For $a, b \in \mathbb{N}_0$ and $m \ge 1$ such that  $a+b\leq m$, define
\begin{equation*}
k^{(a,b)}(r)
=\partial_x^{a}\partial_{x'}^{b}k(x,x').
\end{equation*}
Then
\begin{equation*}
k^{(a,b)}(r) =(-1)^{b} \partial_r^{a+b}k(r).  
\end{equation*}
\end{proposition}

The conditions for this general derivative function $k^{(a,b)}(x, x')$ to be a covariance functions is thus follows: 
\begin{proposition}[Conditions for $k^{(a,b)}$ to be a covariance kernel]
\label{prop:psd-derivative-kernel-main-text}
Assume the kernel $k$ is stationary and isotropic with spectral density $S_k(\omega)$ and that $a+b\le m$. Then the derivative function $k_{a,b}$ is a positive semidefinite covariance kernel if and only if the following set of conditions hold:
\begin{enumerate}
    \item[(1)] $a+b$ is even. 
    \item[(2)] $a,b$ are such that $a+3b\equiv 0\pmod 4$. 
\end{enumerate}

In particular, $a=b$ always satisfies these conditions, so $k^{(a,a)}$ is a covariance kernel.
\end{proposition}
While we show the case of uni-dimensional input $x$, our propositions hold for $p$-dimensional inputs $\bx \in \mathbb{R}^p$. 
We present the proofs in Supplementary Material A.

For a SE covariance function the derivative covariance structure is thus obtained as
\begin{equation} \label{eqn-deriv-se}
    \begin{split}
        k(x_i, x_j) &= \alpha^2 \exp \left(-\frac{(x_i - x_j)^2}{2\rho^2}\right),\\
         k^{(1,0)}(x_i, x_j) &= \alpha^2 \frac{(x_i - x_j)}{\rho^2} \exp \left(-\frac{(x_i - x_j)^2}{2\rho^2}\right),\\
        k^{(1,1)}(x_i, x_j) &= \frac{\alpha^2}{\rho^4}(\rho^2 - (x_i - x_j)^2) \exp\left(-\frac{(x_i - x_j)^2}{2\rho^2}\right).
    \end{split}
\end{equation}
where $\rho > 0$ and $\alpha>0$ are the length-scale and GP marginal SD respectively. 
Further details on the derivative covariance structure and other derivative Mat\'{e}rn covariance functions can be found in \cite{mukherjee_dgp-lvm_2025}.

\subsection{Partial composite Gaussian processes} \label{sec-part-gps}
In real world applications, the exact relationship between $\Bf$ and $\bg$ often remains unknown.
As previously mentioned, under the composite GP framework, it becomes challenging to define the functional forms of $k_{fg}$ and $k_{gf}$ (see Section \ref{sec-comp-gp}) thus affecting the overall choice of the covariance function structure. 
In cases like the derivative GPs where the forms of $k_{fg}$ and $k_{gf}$ are known through $k^{(1,0)}$ and $k^{(0,1)}$ respectively, we face limitations in approximating the model. 
Under the Hilbert space framework, we would require all four quadrants of the joint covariance matrix to satisfy the conditions of Proposition \ref{prop:psd-derivative-kernel-main-text} for a block-wise approximation strategy. 
While the diagonal functions indeed satisfy these conditions, the functions in the off-diagonals do not (see Supplementary Material A).  

In such cases, for practical reasons, we assume $\Bf$ and $\bg$ are independent, thus relaxing the overall composite covariance structure. We refer to the composite GPs with such a covariance structure as partial composite GPs or pcGPs.
The pcGP is therefore modified from Eq.\ref{eqn-comp-gp} as
\begin{equation} \label{eqn-ind-comp-gp}
    \left(\begin{matrix}
        f(x) \\
        g(x)
    \end{matrix}\right)
    \sim
    \mathcal{GP}\left(
    \left(\begin{matrix}
        m_f \\
        m_g
    \end{matrix}\right),
    \left(\begin{matrix}
        k_f & 0\\
        0 & k_g
    \end{matrix}\right)\right),
\end{equation}
where the off-diagonals in the joint covariance function indicates independence between $\Bf$ and $\bg$. 
The rest of the model remains unaltered.
We will focus only on this pcGP structure in this paper and the above notations will be extended to multi-output and latent variable GPs in Section \ref{sec-latent-multi}. 
The properties of the two GPs depend on the choice of the corresponding covariance functions.
As a specific case of pcGPs, we thus introduce partial derivative GPs (pdGPs) that are similarly modified from Eq.\ref{eqn-deriv-gp}. 
Concretely, we arrive at the pdGP formulation by considering $\Bf$ along with the covariance function $k$ followed by substituting $g(x) = \df(x)$ and subsequently $k_g = k^{(1,1)}$ in Eq.\ref{eqn-ind-comp-gp}.

\subsection{Partial composite Hilbert space Gaussian processes } \label{sec-hs-approx}
Exact GPs are known to have a computational complexity of $O(N^3)$ where $N$ is the sample size, thus prohibiting applications on cases where $N$ is large. 
We overcome this scalability issue through HSGP methods \citep{solin_hilbert_2020, riutort-mayol_approx_gps_2022, mukherjee2025hilbert}, where we approximate the covariance function with its spectral density. Briefly, following Bochner's theorem \citep{gihman_linear_2004} and Wiener-Khintchine theorem \citep{AnalTimeSeriesChatfield}, any stationary covariance function $k (r)$ with $r = x - x'$ can be written as 
\begin{equation} \label{eqn-cov-specdens-general}
    S_k(\omega) = \int_{\mathbb{R}}\text{exp}(-i\omega r) k(r) dr
\end{equation}
where $\omega \in \mathbb{R}$ are the inputs in the frequency domain.

Thus, the spectral density for a SE covariance function \citep{gihman_linear_2004} is given as
\begin{equation} \label{eqn-specdens-se}
    S_{k_f}(\omega) = \sqrt{2\pi} \alpha_f^2\rho_f\exp\left(-\frac{1}{2}\rho_f^2 \omega^2 \right)
\end{equation}
where $\rho_f$ and $\alpha_f$ are the covariance function hyperparameters for a GP $f$. 

The HSGP procedure starts by defining a closed set $\Omega \subset [-L, L] \subset \mathbb{R}$ that contains the vector of inputs $\bx$. 
We then write any stationary covariance function with $x, x' \in \Omega$ as
\begin{equation} \label{eqn-covfn-specdens}
    k(x, x') = \sum_{j = 1}^{\infty}S(\sqrt{\lambda_j})\phi_j(x)\phi_j(x'),
\end{equation}
where $S$ is the spectral density of the covariance function $k$. 
Following the Dirichlet conditions \citep{solin_hilbert_2020}, the sets of eigenvalues $\{\lambda_j\}_{j=1}^{\infty}$ and eigenfunctions $\{\phi_j\}_{j = 1}^{\infty}$ are given by
\begin{equation} \label{eqn-eigen-solns}
    \begin{aligned}
        \lambda_j &= \left(\frac{j\pi}{2L}\right)^2, &\\
        \phi_j(x) &= \sqrt{\frac{1}{L}} \sin\left(\sqrt{\lambda_j}(x + L)\right).
    \end{aligned}
\end{equation}
The eigenvalues and eigenfunctions are independent of the choice of covariance functions, and any effects of the hyperparameters for a chosen covariance function are informed through the corresponding spectral density. 
We approximate the covariance function using a linear combination of its basis functions along with the spectral density, eigenvalues and eigenfunctions. 
Considering the first $M$ number of basis terms, we obtain from Eq. \ref{eqn-covfn-specdens}
\begin{equation} \label{eqn-covfn-approx}
    k(x, x') \approx \sum_{j = 1}^{M}S(\sqrt{\lambda_j})\phi_j(x)\phi_j(x').
\end{equation}

In case of the pcGPs with $\Bf$ and $\bg$ having covariance functions $k_f$ and $k_g$, we first derive their corresponding spectral densities $S_{k_f}$ and $S_{k_g}$. 
We then apply the Hilbert space approximation such that 
\begin{equation} \label{eqn-comp-hsgp}
    \left(\begin{matrix}
        f(x) \\
        g(x)
    \end{matrix}\right)
    \sim
    \mathcal{GP}\left(
    \left(\begin{matrix}
        m_f \\
        m_g
    \end{matrix}\right),
    \left(\begin{matrix}
        \tilde{k}_f & 0\\
        0 & \tilde{k}_g
    \end{matrix}\right)\right)
\end{equation}
where $\tilde{k}_f = \sum_{j = 1}^{M}S_{k_f}(\sqrt{\lambda_j})\phi_j(x)\phi_j(x')$ and $\tilde{k}_g = \sum_{j = 1}^{M}S_{k_g}(\sqrt{\lambda_j})\phi_j(x)\phi_j(x')$ are the Hilbert space approximated covariance functions corresponding to $k_f$ and $k_g$ respectively. 
Note that the input space is shared between $f$ and $g$ and thus the eigenvalues $\lambda$ and eigenfunctions $\phi$ are evaluated on the same set of inputs $x$. 
So, any difference in the covariance functions between $f$ and $g$ (different hyperparameters in our case) are, again, only informed through their corresponding spectral densities. 

The full covariance matrices $\mathbf{K}_f$ and $\mathbf{K}_g$ that are generated from $k_f$ and $k_g$ respectively for inputs $x_i$, $i \in \{1, \dots, N\}$ are approximated by its finite basis function as
\begin{equation} \label{eqn-covmat-eigendecomp}
    \mathbf{K}_f \approx \mathbf{\Phi}\mathbf{\Delta}_f\mathbf{\Phi}^T \quad \text{and}\quad \mathbf{K}_g \approx \mathbf{\Phi}\mathbf{\Delta}_g\mathbf{\Phi}^T,
\end{equation}
where $\mathbf{\Delta}_f = \text{diag}(S_{k_f}(\sqrt{\lambda_1}),\ldots,S_{k_f}(\sqrt{\lambda_N}))$, $\mathbf{\Delta}_g = \text{diag}(S_{k_g}(\sqrt{\lambda_1}),\ldots,S_{k_g}(\sqrt{\lambda_N}))$ and $\mathbf{\Phi} \in \mathbb{R}^{N\times M}$ is the matrix of eigenfunctions
\begin{equation} \label{eqn-eigenmatrix}
    \mathbf{\Phi} = \mathbf{\Phi}(\bx) = \begin{bmatrix}
    \phi_1(x_1) & \dots & \phi_M(x_1) \\
    \vdots & \ddots & \vdots \\
    \phi_1(x_N) & \dots & \phi_M(x_N),
    \end{bmatrix}
\end{equation}
for a vector of input points $\bx = (x_1, \ldots, x_N)$. 
We then replace the covariance matrix $\mathbf{K}_f$ and $\mathbf{K}_g$ by its basis approximated eigenvalue decompositions and specify the GPs $\Bf$ and $\bg$ following the linear representation from \cite{riutort-mayol_approx_gps_2022, mukherjee2025hilbert} as
\begin{equation} \label{eqn-gp-approx}
\begin{aligned}
     f(\mathbf{x}) &\approx \mu_f + \sum^{M}_{j = 1} \left(S_{k_f}(\sqrt{\lambda_j})\right)^{1/2} \mathbf{\Phi}(\bx) \beta_{fj}, \\
      g(\mathbf{x}) &\approx \mu_g + \sum^{M}_{j = 1} \left(S_{k_g}(\sqrt{\lambda_j})\right)^{1/2} \mathbf{\Phi}(\bx) \beta_{gj},
\end{aligned}
\end{equation}
where $\beta_{fj} \sim \mathcal{N}(0, 1)$ and $\beta_{gj} \sim \mathcal{N}(0, 1)$ for $j \in \{1,\ldots,M\}$. 
This results pcHSGPs with $\Bf$ and $\bg$ having eigen-decomposed approximate covariance functions. 
Using pcHSGP, we then simultaneously model outputs $\by_f$ and $\by_g$ with a shared input $\bx$ as shown in Eq.\ref{eqn-comp-gp-model2}.

\subsection{Partial derivative Hilbert space Gaussian processes} \label{sec-deriv-hsgp}
Under the Hilbert space methods, we approximate the covariance functions of the GP $\Bf$ and its derivative process $\bdf$ with its spectral density. 
Based on the Proposition \ref{prop:derivative-kernel-main-text}, related to the general structure of derivative covariance functions (Section \ref{sec-deriv-gp}), we state the following:
\begin{proposition}[Spectral representation of derivative kernels]
\label{prop:spectral-derivative-main-text}
Let $k^{(a,b)}(x, x') = k^{(a,b)}(r)$ be a stationary covariance kernel, then its spectral density $S^{(a,b)}(\omega)$ is given by
\[
S^{(a,b)}(\omega)=(i\omega)^{a} (-i\omega)^{b} S_k(\omega).
\]
\end{proposition}
See Supplementary Material A for the proof. Additionally, we derive the spectral densities and the conditions under which they satisfy Proposition \ref{prop:spectral-derivative-main-text} for SE as well as the general Mat\'{e}rn family of covariance functions (see Corollary 1 and 2 in Supplementary Material A). Furthermore, our choice of $k$ preserves the assumptions of $k^{(1,1)}$ being isotropic and subsequently $S_{k^{(1,1)}}$ having an analytic functional form, thus satisfying the Hilbert space approximation conditions \citep{solin_hilbert_2020}.

Taking the example of the derivative SE covariance function $k^{(1,1)}$ shown in Section \ref{sec-deriv-gp}, we thus obtain the spectral density of the using Proposition \ref{prop:spectral-derivative-main-text}:
\begin{equation} \label{eqn-specdens-dse}
    S_{k^{(1,1)}}(\omega) = \omega^2 S_k(\omega) =\sqrt{2\pi}\omega^2\alpha^2\rho\exp\left(-\frac{1}{2}\rho^2\omega^2\right),
\end{equation}
where $S_k(\omega)$ is the spectral density of the standard SE covariance function (as seen in Eq.\ref{eqn-specdens-se}).

We resort again to the pcGP structure for reasons described in Section \ref{sec-part-gps}. 
Thus, the derivative HSGP is now specified as
\begin{equation} \label{eqn-deriv-hsgp}
    \left(\begin{matrix}
        f(x) \\
        g(x)
    \end{matrix}\right)
    \sim
    \mathcal{GP}\left(
    \left(\begin{matrix}
        m_f \\
        m_g
    \end{matrix}\right),
    \left(\begin{matrix}
        \tilde{k} & 0\\
        0 & \tilde{k}^{(1,1)}
    \end{matrix}\right)\right)
\end{equation}
such that $\tilde{k} = \sum_{j = 1}^{M}S_k(\sqrt{\lambda_j})\phi_j(x)\phi_j(x')$ and $\tilde{k}^{(1,1)} = \sum_{j = 1}^{M}S_{k^{(1,1)}}(\sqrt{\lambda_j})\phi_j(x)\phi_j(x')$ are the Hilbert space approximated covariance functions corresponding to $k$ and $k^{(1,1)}$ respectively. 
The rest of the steps for the approximation procedure remains the same. 
Using pdHSGPs, we overcome the practical limitations of derivative GPs outlined in \cite{mukherjee_dgp-lvm_2025}. 
However, in pdHSGPs, we don't model and learn the functional relationship between $\Bf$ and $\bdf$ through the first order partial derivatives of the covariance functions. 
Thus, pdHSGPs are, to some degree, misspecified in cases where the underlying data generating process is a full derivative GP model. We show the degree of trade-off for this misspecification as a means to a scalable solution for derivative GPs in Section \ref{sec-sim-study}.

\subsection{Extending the partial composite structure} \label{sec-latent-multi}
We extend the pcHSGPs and pdHSGPs to model multi-dimensional outputs and latent variable inputs. 
In this paper, we follow a similar framework pertaining to the multi-output latent variable models considered in \cite{mukherjee_dgp-lvm_2025, mukherjee2025hilbert}. 

\subsubsection{Multi-dimensional outputs}\label{sec-multi-hsgps}
We specify the multi-output version of the composite HSGPs with response variables $(\by_{f_1},\dots,\by_{f_D})$ and $(\by_{g_1},\dots,\by_{g_D})$ over $D>1$ output dimensions \citep{gp_rasmussen_williams_2006}, using $y_{f_{di}}$ and $y_{g_{di}}$ to denote the response for $d^{th}$ dimension and $i^{th}$ sample. 
As the usual approach, we first set up $D$ independent, univariate Gaussian processes $\Bf_d$ and $\bg_d$ each with their own set of hyperparameters $\bm{\theta}_{f_d}$ and $\bm{\theta}_{g_d}$ \citep{teh_semiparametric_2005,mukherjee_dgp-lvm_2025, mukherjee2025hilbert}. 
We extend the composite HSGPs to multi-output GPs by modifying Eq. \ref{eqn-gp-approx} such that
\begin{equation} \label{eqn-gp-approx-multi}
\begin{aligned}
    f_d(\mathbf{x}) &\approx \mu_{f_d} \sum^{M}_{j = 1} \left(S_{k_{f_d}}(\sqrt{\lambda_j})\right)^{1/2} \mathbf{\Phi}(\bx) \beta_{f_{jd}},\\
    g_d(\mathbf{x}) &\approx \mu_{g_d} + \sum^{M}_{j = 1} \left(S_{k_{g_d}}(\sqrt{\lambda_j})\right)^{1/2} \mathbf{\Phi}(\bx) \beta_{g_{jd}},
\end{aligned}
\end{equation} 
where $\beta_{f_{jd}} \sim \mathcal{N}(0, 1)$ and $\beta_{g_{jd}} \sim \mathcal{N}(0, 1)$ for the corresponding GP approximations. 
The dimensions of $\mathbf{\Phi}$ thus remains same, since it depends only on sample size $N$ and number of basis functions $M$. With the multi-dimensional outputs $D$, the exact GPs have a computational complexity of $O(N^3D + ND^2)$ where $N$ is the sample size of outputs. 
Through this HSGP formulation using $M$ number of basis functions, we sharply decrease it to $O(NMD+ND^2)$. 

The univariate GPs are then related to one another by linearly combining them with a ($D$-dimensional) across-dimension uniform correlation matrix $\mathbf{C}_f$ and $\mathbf{C}_g$ \citep{teh_semiparametric_2005, bonilla_multi-task_2007}. 
Specifically, for each $i^{th}$ sample, we obtain a vector of across-dimension correlated GP values as
\begin{equation} \label{multi-gps}
\left(\begin{array}{r} f^*_1(x_i) \\ \dots \\ f^*_D(x_i) \end{array}\right) 
= \mathbf{A}_f \times \left(\begin{array}{r} f_1(x_i) \\ \dots \\ f_D(x_i) \end{array}\right) 
\quad \text{and} \quad 
\left(\begin{array}{c} g^*_1(x_i) \\ \dots \\ g^*_D(x_i) \end{array}\right) 
= \mathbf{A}_g \times \left(\begin{array}{c} g_1(x_i) \\ \dots \\ g_D(x_i) \end{array}\right),
\end{equation}
where $\mathbf{A}_f$ and $\mathbf{A}_g$ are the Cholesky factors of $\mathbf{C}_f$ and $\mathbf{C}_g$ respectively such that $\mathbf{C}_f = \mathbf{A}_f \mathbf{A}_f^T$ and $\mathbf{C}_g = \mathbf{A}_g \mathbf{A}_g^T$. 
This way, multi-output GPs combine two dependency structures, one within dimensions (and across observations) as expressed by the univariate GPs through corresponding covariance functions and one across output dimensions (but within observations) as expressed by $\mathbf{C}_f$ and $\mathbf{C}_g$ for both the GPs. 
Adding independent Gaussian noise to our derivative multi-output GP model, we extend Eq.\ref{eqn-comp-gp-model2} for all $d$ and $i$:
\begin{equation} \label{multi-GPs}
y_{f_{di}} \mid f^*_d \sim \mathcal{N}(f^*_d(x_i), \sigma_{f_d}^2) \quad \text{and} \quad y_{g_{di}} \mid g^*_d \sim \mathcal{N}(g^*_d(x_i), \sigma_{g_d}^2).\quad 
\end{equation}
The structure remains same for the derivative version of the composite HSGP where we again consider $g(x) = \df(x)$ along with the corresponding changes in the covariance function (and spectral densities).

\subsubsection{Latent variable inputs}\label{sec-latent-hsgps}
Within latent-variable composite HSGPs, a single input vector $\mathbf{x}$ shared between $\Bf$ and $\bg$ (and $\bdf$ in case of derivative HSGPs) is considered as unobserved and treated similar to other estimable parameters. 
To achieve this, we first consider an observed quantity $\tilde{\bx}$ that acts as a noisy measurement to the latent inputs $\bx$. 
From a Bayesian perspective, $\tilde{\bx}$ is a quantity based on which we define a prior for the latent $\bx$, which is then refined by the HSGPs learning from $\by_f$ and $\by_g$ (and $\bdy$ for derivative HSGPs). 
Specifically, we assume that the measurements $\tilde{\bx}$ are Gaussian with known measurement SD $s$ such that we can write for each observation $i$:
\begin{equation} \label{eqn-latent prior}
    \tilde{x}_i \sim \mathcal{N}(x_i, s^2).
\end{equation}

The vector of latent inputs $\bx$ is then passed on to the approximation step in Eq.\ref{eqn-gp-approx} (and subsequently Eq.\ref{eqn-gp-approx-multi} for the multi-output case). 
We call the resulting model (multi-output) latent variable HSGPs. 
Latent variable GPs, exact or approximate, are more difficult to fit than their manifest counterparts. 
The primary reasons are the substantial increase in the number of estimable parameters as well as identification issues arising due to both $\bx$ and $\rho$ now being treated as unknown. 
More details on these issues are discussed in \cite{mukherjee_dgp-lvm_2025, mukherjee2025hilbert}. 
Briefly, we overcome these issues by pooling information through multiple output dimensions for a single latent $\bx$. 
The noisy measurements $\tilde{\bx}$ additionally help identifying the length-scale $\rho$ and latent $\bx$. 

\section{Simulation study}
\label{sec-sim-study}

For latent variable models, we lack ground truth values in real-world scenarios and thus it is challenging to validate such models based on real data alone. 
We therefore design simulation studies where we validate our developed methods against simulated ground truths. 
We compare pcHSGPs and pdHSGPs to their exact counterparts on their latent variable estimation accuracy and inference speed. 
The overall simulation study design is inspired from \cite{mukherjee_dgp-lvm_2025, mukherjee2025hilbert}, but extended to include the pcHSGPs, pdHSGPs, and their exact models pcGP and pdGP respectively.
We fit all the models involved using full Bayesian inference via MCMC sampling in Stan \citep{Stan_guide_2024}. 
The details of the model inference procedure are presented in Supplementary Material B.

\subsection{Data generating process}
\label{sec-sim-data}

We consider two data generating processes as our simulation scenarios. 
In the first scenario, we generate the data from an exact pcGP (from Eq.\ref{eqn-ind-comp-gp}). 
Under this scenario, we showcase the advantages of using the approximate pcHSGP model when the true underlying data is the exact pcGP. 
In the second scenario, we generate data using a full derivative GP (dGP) (see Eq.\ref{eqn-deriv-gp}). 
For the dGP data scenario, we compare the levels of model misspecification based on two counts: partial covariance structure (via pdGPs and pdHSGPs) and covariance function mismatch (via pcGPs and pcHSGPs).

In the above scenarios, we have a sample size of $N = 20$ since exact GPs are considered among the set of models to be compared. 
Thus, as third scenario, we consider dGP data with higher $N = 100$ where we only compare different HSGPs.   
Under this scenario, in addition to pcHSGP and pdHSGPs, we consider single HSGPs (sHSGP) and single derivative HSGP (sdHSGP) where we only have single GP functions $f$ and $\df$ respectively. 
Including the sHSGP and sdHSGP answers the question regarding the advantages of composite data sources versus a simpler single source of information. 
Although HSGPs could easily handle much higher $N$ as we show in our case study (in Section \ref{sec-case-study}), we use only use $N = 100$ to reduce the overall runtime of the simulations. An outline of our simulation scenarios are presented in Table \ref{tab:sim-study}.
 \begin{table}[!ht]
     \centering
     \caption{Simulation study design}
     \begin{tabular}{c c c c}
     \toprule
      Simulation scenario & Sample size ($N$) & Data generating process & Covariance function  \\
  \midrule
 1 & 20 & pcGP & partial composite SE \\ 
 2 & 20 & dGP & joint derivative SE \\
 3 & 100 & dGP & joint derivative SE \\
  \bottomrule
     \end{tabular}
     \begin{tablenotes}
        \item \small \textit{Note: All simulation scenarios consists of output dimensions $D = 5, 10 \text{ } \text{and} \text{ } 20$. 
        The data generating processes are partial composite GP (pcGP) and joint derivative GP (dGP). The covariance functions are based on Squared Exponential (SE) function.}
     \end{tablenotes}
     \label{tab:sim-study}
 \end{table} 
 
Our data generating conditions are specified in a way that ensures a fair amount of non-linearity (via length-scales) as well as a good signal-to-noise ratio (via marginal and error SDs) in the simulated data.
Concretely, for the pcGP data generating process, we sample multi-dimensional length-scales $\brho_f \sim \text{Normal}^+(1, 0.05^2)$ and $\brho_g \sim \text{Normal}^+(0.7, 0.05^2)$, GP marginal SDs $\balpha_f \sim \text{Normal}^+(3, 0.25^2)$ and $\balpha_g \sim \text{Normal}^+(2, 0.25^2)$, and error SDs $\bsigma_f \sim \text{Normal}^+(1, 0.25^2)$ and $\bsigma_g \sim \text{Normal}^+(0.75, 0.25^2)$. 
In case of the dGP data, we first induce a scale difference between the outputs $\by$ and derivatives $\bdy$ (as well as $\Bf$ and $\bdf$ subsequently) using a scaling proportion $\lambda = 10$. 
Thus, we re-purpose the GP marginal SDs and error SDs of composite GPs to be $\balpha_f = \balpha \, , \balpha_g = \bdalpha$ and $\bsigma_f = \bsigma \,, \bsigma_g = \bdsigma$ to remain coherent with the derivative GP notations. We induce this scale difference by first sampling marginal and error SDs for $\bdf$ and $\bdy$ where $\bdalpha \sim \text{Normal}^+(3, 0.25^2)$ and $\bdsigma \sim \text{Normal}^+(1, 0.25^2)$. 
Then for the outputs $\by$ (and functions $\Bf$), we set $\balpha = \lambda\bdalpha$ and $\bsigma = \lambda\bdsigma$. Under this data generating process, we only have a single length-scale $\brho \sim \text{Normal}^+(1, 0.05^2)$. More details regarding the implications and relevance of this scaling proportion $\lambda$ can be found in \cite{mukherjee_dgp-lvm_2025}. 

In pcGP data scenario, we sample across-output dimension correlation matrices $\mathbf{C}_f$ and $\mathbf{C}_g$ from the $\text{LKJ} (\eta = 1)$ distribution \citep{LKJcholesky2009}. A value of $\eta = 1$ implies $\mathbf{C}_f$ and $\mathbf{C}_g$ to be uniformly distributed within the set of all correlation matrices of dimension $D$. 
For the dGP data scenario, we only have a single large between-dimension (output) correlation matrix $\mathbf{C} \sim \text{LKJ}(\eta = 1)$. 
For both data generating processes, we sample the constant mean functions $m_f$ and $m_g$ (which is $m_{\df}$ for dGP data) from $\text{Normal}^+(0, 5^2)$. 
We generate the ground truth latent inputs for the simulation scenarios as $x_i \sim \text{Uniform}(0, 10)$, such that $i = 1, \ldots, N$. 
Furthermore, we assume a prior measurement SD of the noisy $\tilde{\bx}$ as $s = 0.3$ (see Section \ref{sec-latent-multi}). 
For the number of output dimensions $D$, we consider the cases $D = 5, 10 \text{ } \text{and} \text{ } 20$. 
We perform 50 simulation trials for each of the choices of $D$ for every simulation scenario. 

\subsection{Model specifications}
\label{sec-sim-model}

We fit a set of exact GPs and their approximated HSGPs for the simulation scenarios discussed above. 
Under the pcGP data scenario, we only compare the true model against its approximation pcHSGP. In this case, we check the benefits of using Hilbert space approximations for composite GP models.
For the dGP data scenario, we compare pcGP and pdGP along with their approximations pcHSGP and pdHSGP respectively. In the third scenario with dGP data and $N = 100$ we only involve HSGPs. In this case, we include sHSGP and sdHSGP having a single source of information in addition to pcHSGPs and pdHSGPs. Including sHSGPs and sdHSGPs checks the advantages of composite models with two sources of data as opposed to simpler HSGPs with a single data source. The entire set of comparative models along with their features are presented in Table \ref{tab:model-comp}.

 \begin{table}[!ht]
     \centering
     \caption{GP model specifications involved in our simulation study}
     \begin{tabular}{c c c c c c}
     \toprule
      Model name & Model type & Simulation scenarios & Composite & Derivative & Partial \\
  \midrule
 pcGP & E & 1, 2  & \cmark & \xmark & \cmark \\ 
 dGP & E & 2  & \cmark & \cmark & \xmark \\
 pdGP & E & 2 k & \cmark & \cmark & \cmark \\
 pcHSGP & A & 1, 2, 3  & \cmark & \xmark & \cmark \\
 pdHSGP & A & 2, 3  & \cmark & \cmark & \cmark \\
 sHSGP & A & 3 k & \xmark & \xmark & \xmark \\
 sdHSGP & A & 3  & \xmark & \cmark & \xmark \\
  \bottomrule
     \end{tabular}
     \begin{tablenotes}
        \item \small \textit{Note: Model names denote partial composite GPs (pcGP) and partial derivative GPs (pdGP) and their approximations pcHSGP and pdHSGP respectively. Additionally, we have the joint derivative GP (dGP), single HSGP (sHSGP) and single derivative HSGP (sdHSGP). The models are either exact (E) or approximate (A) in nature. We use the simulation scenario numbering from Table \ref{tab:sim-study}.}
     \end{tablenotes}
     \label{tab:model-comp}
 \end{table} 

The priors for all model parameters are the same as those used in the data generating process, a requirement for the uncertainty calibration tests (shown in Supplementary Material E).
Thus, in the pcGP data simulation scenario, the models pcGP and pcHSGP length scales have multi-dimensional hyperparameters $\brho_f \sim \text{Normal}^+(1, 0.05^2)$ and $\brho_g \sim \text{Normal}^+(0.7, 0.05^2)$, GP marginal SDs $\balpha_f \sim \text{Normal}^+(3, 0.25^2)$ and $\balpha_g \sim \text{Normal}^+(2, 0.25^2)$, and error SDs $\bsigma_f \sim \text{Normal}^+(1, 0.25^2)$ and $\bsigma_g \sim \text{Normal}^+(0.75, 0.25^2)$. 
Under the dGP data scenario, the pdGP and pdHSGP involving derivative SE structure only have a single length scale $\brho \sim \text{Normal}^+(1, 0.05^2)$. 
The marginal SDs $\balpha \sim \text{Normal}^+(30, 2.5^2)$ and $\bdalpha \sim \text{Normal}^+(3, 0.25^2)$. 
The error SDs are specified to follow $\bsigma \sim \text{Normal}^+(10, 2.5^2)$ and $\bdsigma \sim \text{Normal}^+(1, 0.25^2)$. 
Under this scenario, in case of pcGP and pcHSGP, the priors for both length-scales $\brho_f$ and $\brho_g$ remain the same as $\brho$. 
The marginal $\balpha_f, \, \balpha_g$ and error SDs $\bsigma_f, \, \bsigma_g$ follow the same prior distributions as $\balpha, \, \bdalpha$ and $\bsigma, \, \bdsigma$ respectively. 
For the dGP data scenario with $N = 100$, we have the sHSGP and sdHSGP, having only a single covariance structure, involves a single set of hyperparameters $\brho, \, \balpha, \, \bsigma$ and $\brho, \, \bdalpha, \, \bdsigma$ respectively. 
The respective priors for these hyperparameters remain the same as described above.

Based on the suggestions in \cite{riutort-mayol_approx_gps_2022, mukherjee2025hilbert}, we select the boundary conditions $L$ for all the HSGP models involved, by multiplying a scaler adjustment $c = 1.25$ to the range of the input prior $\tilde{\bx}$. 
This way, we ensure that we have a closed interval $[-L, L]$ that prevents input values $\bx$ to be near the boundaries. 
While there is an empirical relation that determines the minimum number of required basis functions $M$ \citep{riutort-mayol_approx_gps_2022}, it is only applicable for standard Mat\'{e}rn class of covariance functions. 
Since it has not been extended for composite (and derivative) covariance functions, we select $M = 30$ for the all the simulation scenarios based on the results of \cite{mukherjee2025hilbert}, which is enough to capture the non-linearity for both SE and derivative SE functions through their spectral approximations.

We implemented all models in Stan \citep{Stan_guide_2024} and conducted the simulation studies with the \texttt{rstan} interface \citep{Rstan_2023}. 
The models are fitted with a single MCMC chain of 2000 iterations of which the first 1000 are discarded as warm-up. 
\cite{mukherjee_dgp-lvm_2025} show that model convergence is similar for multiple chains when fitting latent variable GPs for the SE and derivative SE covariance functions. 
Thus, we run only a single chain per model to parallelize over the 50 trials to reduce overall computation times. 

\subsection{Latent variable estimates}\label{sec-latent-est}
We evaluate the latent variable estimation accuracy for our proposed models under all simulation scenarios. 
To this end, we compare posterior samples of latent $\bx$ denoted by $x_{post}$ from our fitted models to their true values $\bx_{true}$ using $\text{RMSE}(\bx_{post}) = \sqrt{\mathbb{E}\left((\bx_{post} - \bx_{true})^2\right)}$. 
Under the different simulation conditions described before, we study the effects of our various model choices on $\text{RMSE}(\bx_{post})$. We summarize these effects through a multilevel model setup described in Supplementary Material D. 
Our results show the HSGPs out-perform their exact GP counterparts in terms of RMSE, for latent variable estimation, in all simulation scenarios. 
\begin{figure}[!ht]
    \centering
    \includegraphics[width = \linewidth]{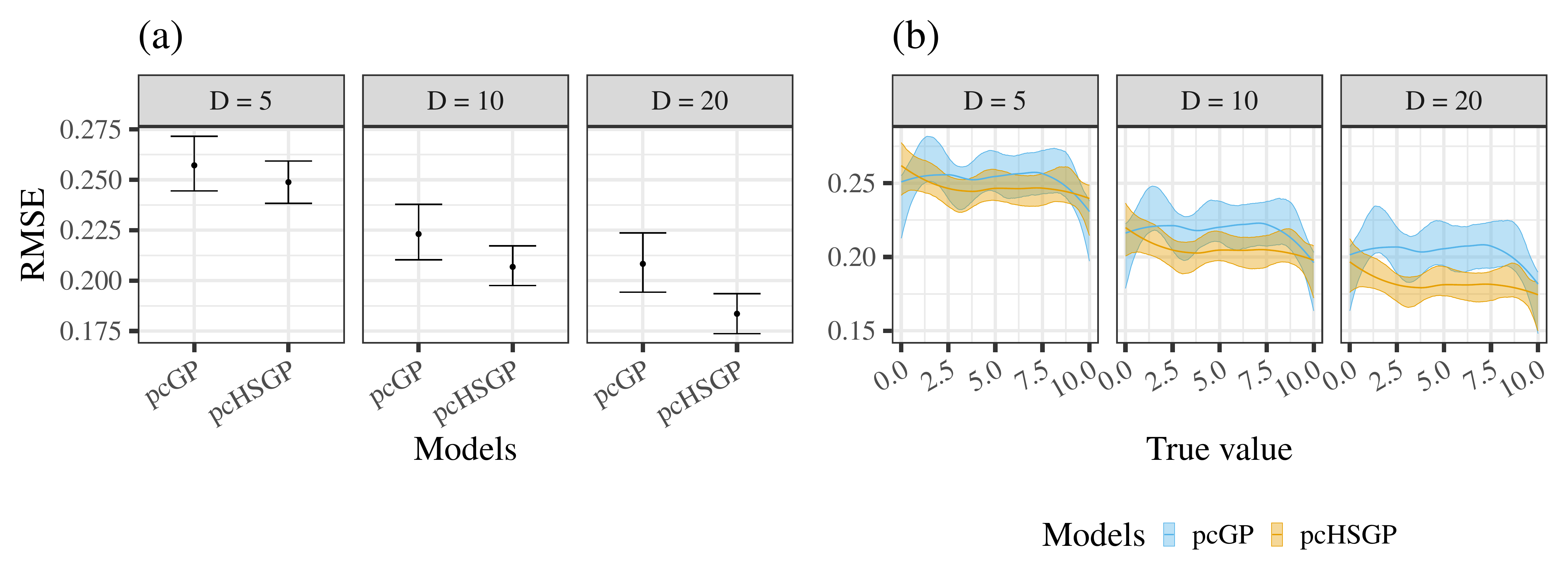}
    \caption{\textit{pcGP data scenario: a) RMSE on recovery of latent inputs for the data generating process pcGP and its approximation pcHSGP. b) Model performance in terms of RMSE for recovering various true values across the input space. Figures are shown for output dimensions $D = 5, 10$ and $20$.}}
    \label{fig:pcgp-data-latentx}
\end{figure}
Specifically, for the pcGP data scenario (Fig. \ref{fig:pcgp-data-latentx}), we see up to 14\% decrease in RMSE (in case of higher output dimensions) with pcHSGPs when compared to the exact pcGP (true data generating process) while also drastically speeding up model fitting.

\begin{figure}[!ht]
    \centering
    \includegraphics[width = \linewidth]{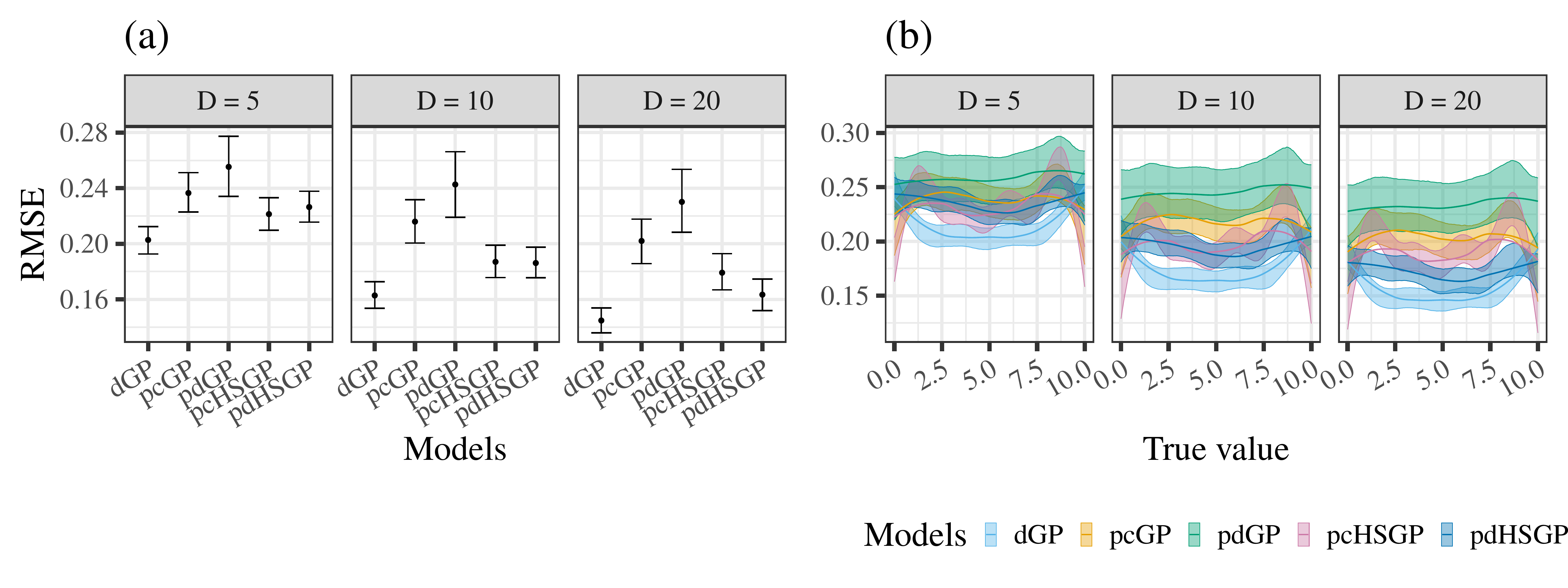}
    \caption{\textit{dGP data scenario: a) RMSE on recovery of latent inputs for all fitted models. Models pcGP and pcHSGP refer to partial composite GPs (exact and approximate). pdGP and pdHSGP refers to the exact and approximate partial derivative GPs. b) Model performance in terms of RMSE for recovering various true values across the input space. Figures are shown for output dimensions $D = 5, 10$ and $20$.}}
    \label{fig:dgp-data-all-latentx}
\end{figure}

In Fig. \ref{fig:dgp-data-all-latentx}, we demonstrate the effects of estimating latent variables under several covariance function misspecifications. 
When compared to the ground truth dGP, naturally, we see an increase in overall RMSE for estimating latent $\bx$ for both pcGP and pdGP. 
This is a direct consequence of using a partial covariance structure to model data generated from a joint derivative structure, not to mention a different covariance function with the pcGP. 
However, the respective HSGPs for the partial GPs readily overcomes this by showing a greater accuracy (as compared to their exact counterparts) in estimating latent variable inputs. 
Concretely, the pcHSGPs show an overall decrease in RMSE as compared to the exact pcGP by up to 10\%. As for the pdHSGP, we see an overall decrease of up to 33\% when compared to the exact pdGP model. 
As for estimation speed, for a dataset with $N=20$ and $D = 20$, the full derivative GP took on average 6.05 hrs. The exact pcGPs and pdGPs took 2.03 hrs and 3 hrs respectively. Both the pcHSGPs and pdHSGPs on the other hand only took 0.22 hrs to fit the same dataset
Under this dGP data scenario with $N = 20$, while the pcHSGPs and pdHSGPs do not yet reach the same accuracy as the ground truth model dGP, they have a clear advantage in terms of model fitting speed. 

\begin{figure}[!ht]
    \centering
    \includegraphics[width = \linewidth]{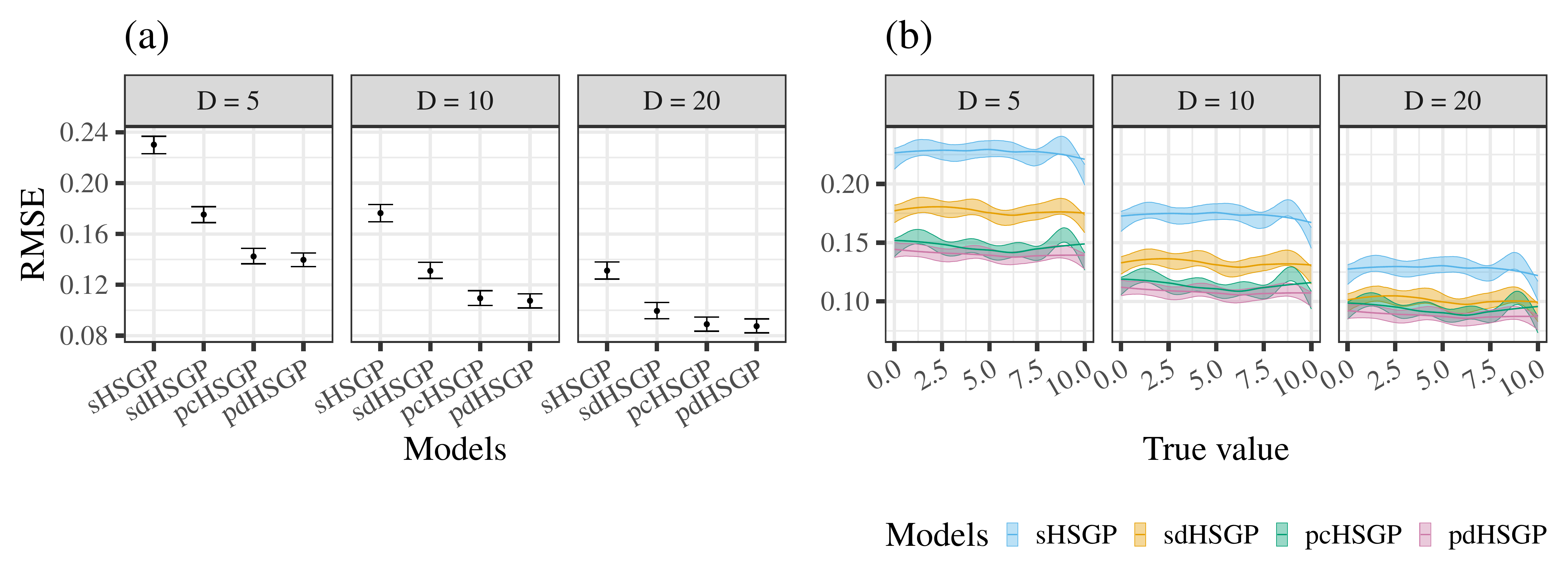}
    \caption{\textit{Derivative GP scenario: a) RMSE on recovery of latent inputs for different Hilbert space GPs. sHSGP and sdHSGPs depict models with either SE or derivative SE covariance function based on a single data source. They are compared the partial composite pcHSGPs and derivative pdHSGP models. b) Model performance in terms of RMSE for recovering various true values across the input space. Figures are shown for output dimensions $D = 5, 10$ and $20$.}}
    \label{fig:dgp-data-hsgps-latentx}
\end{figure}

The results from the dGP data with $N = 100$ scenario (in Fig. \ref{fig:dgp-data-hsgps-latentx}) demonstrates the effects on various choices of HSGPs on latent variable estimation under a larger sample size. 
In this simulation scenario, we additionally considered models with a single covariance function sHSGP and sdHSGP analyzing either the outputs $\by$ or their derivatives $\bdy$ separately. 
We readily see that the composite models pcHSGP and pdHSGP outperforms sHSGP and sdHSGPs (having up to 42\% lower RMSE) in estimating latent $\bx$. 

Through the results of this simulation scenario, we make a couple of important observations. Firstly, among the sHSGP and sdHSGP, the latter outperforms the former in estimating latent $\bx$. 
This is due to the standard SE covariance function being inadequate in modeling the data generated from a much more complex dGP structure. 
Secondly, both pcHSGP and pdHSGP perform almost equally well, under the same data generating condition but with the higher $N = 100$ sample size.
Further, upon comparing the results across Fig. \ref{fig:dgp-data-all-latentx} and Fig. \ref{fig:dgp-data-hsgps-latentx}, we see that pcHSGPs and pdHSGPs obtain a much lower RMSE of up to 43\% (under higher sample size) as compared to ground truth dGP that can only be a feasible modeling option under low sample sizes. 
For $N = 100$ (and $D = 20$), the pcHSGPs and pdHSGPs took 1.26 hrs and 1.15 hrs, respectively. 
Further results including model convergence, uncertainty calibration for latent variable estimates, as well as hyperparameter estimation accuracy are provided in Supplementary Materials C, E and F respectively.

\section{Real-world case study}\label{sec-case-study}
We demonstrate our proposed methods on a real-world single-cell RNA sequencing dataset involving the maturation process of erythroid cells from progenitor cells \citep{pijuan-sala_single-cell_2019}. 
Single-cell RNA sequencing is a technique that allows the measurement of RNA molecule abundance in single cells \citep{haque_practical_2017}. 
Through these measurements, we study the description of the underlying cells on a molecular level. 
By arranging all cells along a latent trajectory, it is possible to uncover and characterize an entire biological process. 
In our case study, the biological process contains two blood progenitors and three maturing erythroid cell stages. 
The observed process describes a trajectory starting from blood progenitors which develop into matured erythroid cells without any bifurcations or cyclical dynamics. 
The dataset contains several experimental time points being the time points of sample collection from original embryonic tissue, which correlate with the maturation process of the cells.
\begin{figure}[!ht]
    \centering
    \includegraphics[width = \linewidth]{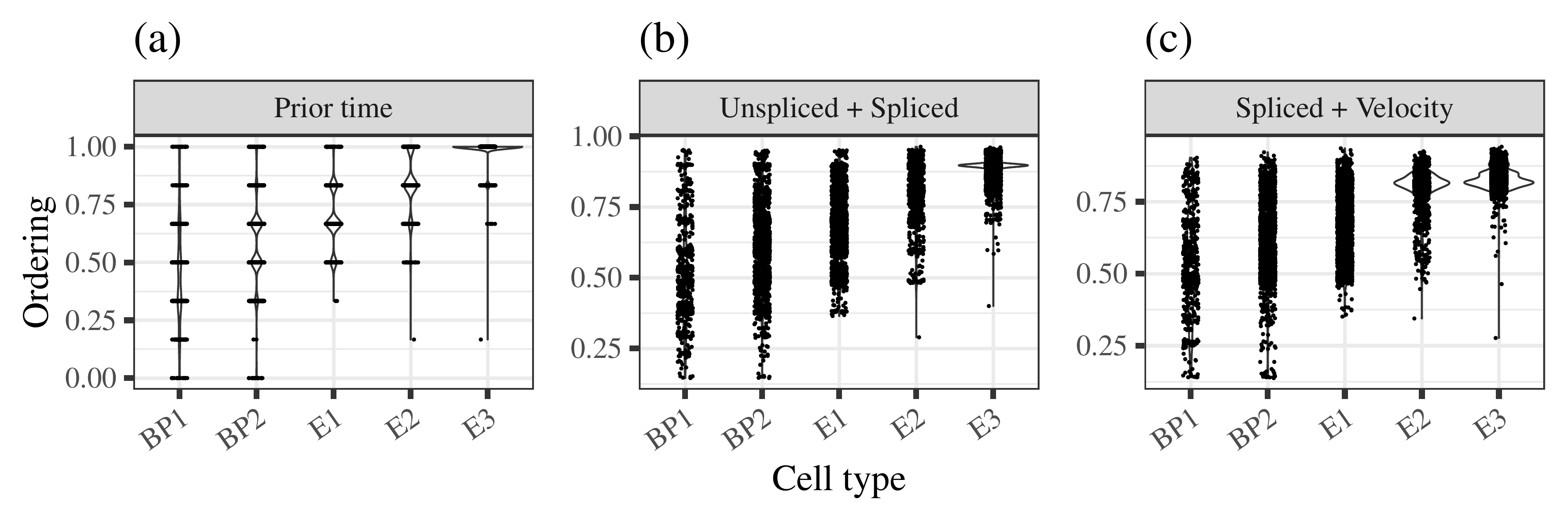}
    \caption{\textit{a) Distribution of discrete experimental time by cell type. b) Posterior latent continuous cell ordering from modeling unspliced and spliced gene expression using pcHSGP. c) Posterior latent continuous cell ordering from modeling spliced gene expression and RNA velocity using pdHSGP. The x-axis denotes the cell types blood progenitors (BP) and erythroid (E).}}
    \label{fig:case-study-cont-latentx}
\end{figure}

The pre-annotated dataset was acquired from the \textit{scvelo} \citep{bergen_generalizing_2020} package and contains $N=9815$ individual cells. 
Based on the suggestions of \cite{barile_coordinated_2021} we use the expression levels of $D=14$ genes that are relevant in understanding this biological process. 
The names of these set of $14$ genes are listed in the Supplementary Material G. 
We specifically demonstrate two different approaches of estimating the latent cellular ordering based on the same biological process. 
As our first approach, we consider two different gene expression levels to estimate latent cellular ordering \citep{hensman_hierarchical_2013, ahmed_grandprix_2019}. 
Specifically, we model the unspliced and spliced gene expressions \citep{trapnell_dynamics_2014} together using our developed pcHSGP to estimate the cellular ordering as latent inputs. 
As an alternative approach, we consider the spliced RNA gene expression levels and RNA velocity \citep{la_manno_rna_2018}, a derivative information that is calculated based on the rate of change between spliced and unspliced gene expression levels. 
In this case, we use pdHSGP to simultaneously model spliced gene expression and their derivative RNA velocity to estimate the same latent cellular ordering. 
As a pre-processing step, we standardize the gene expression values (and RNA velocities) for both the approaches. 
This is done to overcome the high amount of variations among different gene (outputs) thus making it easy to specify priors on covariance function hyperparameters. 
\begin{figure}[!ht]
    \centering
    \includegraphics[width = \linewidth]{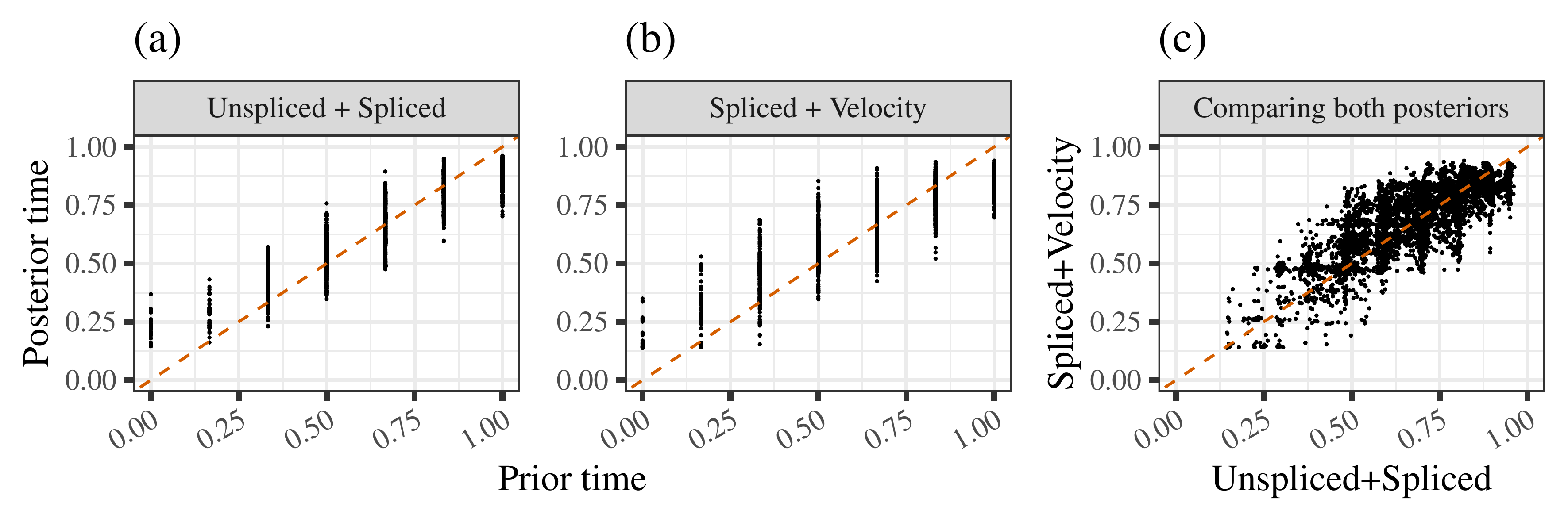}
    \caption{\textit{Deviations of posterior latent cell ordering from discrete experimental times based on a) modeling unspliced and spliced gene expression using pcHSGP and b) modeling spliced gene expression and RNA velocity using pdHSGP. c) Comparison of posterior latent cell orderings from both the models.}}
    \label{fig:case-study-latentx-scatter}
\end{figure}

For the pcHSGP model (used in the first approach), we specify the priors for length-scales $\rho_f, \, \rho_g \sim \text{Normal}^+(0.3, 0.1^2)$, GP marginal SDs $\alpha_f, \, \alpha_g \sim \text{Normal}^+(0.5, 0.1^2)$, and error SDs $\sigma_f, \, \sigma_g \sim \text{Normal}^+(0.5, 0.1^2)$. 
The prior choices for the length-scales reflect similar levels of non-linearity in the functions $\Bf$ and $\bg$ as in our simulation studies. 
The marginal and error SD priors are based on the standardized gene expression values. 
In case of the pdHSGP model, we use the same respective prior distributions for the set of hyperparameters length-scale $\rho$, GP marginal SDs $\alpha, \, \dalpha$ and error SDs $\sigma, \, \dsigma$. 
We assume the experimental time $\tilde{\bx}$ ranging between 0 and 1 as a noisy measurement around $\bx$ with a measurement SD $s = 0.1$ (see Section \ref{sec-latent-hsgps}). 
Given the range of $\tilde{\bx}$ we use a prior of $\bx \sim \text{Uniform}(0, 1)$.  

The HSGPs are specified using the same boundary conditions as in our simulation studies (see Section \ref{sec-sim-model}). 
Choosing an adequate number of basis functions $M$ depends on the range of inputs, the length-scale prior, and the boundary conditions \citep{riutort-mayol_approx_gps_2022, mukherjee2025hilbert}. 
For both our approaches, we use $M = 10$ which is higher than the suggested number of basis functions for a SE covariance function under our model setup. 
We erred on the side of caution with a larger $M$, since the relationship between the minimum number of basis functions for derivative covariance functions needs to be studied further.

We show the estimated cellular ordering using the pcHSGP as well as pdHSGP in Fig. \ref{fig:case-study-cont-latentx}. Our estimated cellular ordering exhibits a continuous temporal sequence based on each cells. 
From a biological perspective, this continuous nature of the ordering is crucial for understanding the trajectory of the biological process, as opposed to only a few discrete points \citep{trapnell_dynamics_2014}. 
We thus enable future research that better describes the underlying transition processes of blood progenitors to erythroid cells based on our estimated continuous cellular ordering.
Secondly, the estimated posteriors shows strong deviations from the prior experimental time for both the approaches in Fig. \ref{fig:case-study-latentx-scatter}(a) and (b). 
Combined with the evidence provided in our simulation studies, we attribute these deviations towards learning the true cellular ordering. 
Furthermore, in Fig. \ref{fig:case-study-latentx-scatter}(c) we see that the estimated orderings from pcHSGP (analyzing unspliced and spliced gene expressions) and pdHSGP (analyzing spliced gene expressions and RNA velocity) deviate from one another indicating independent information to be learned about the biological process from either of the approaches.

\section{Discussion}\label{sec-discuss}
In this paper, we develop a scalable class of models for latent variable estimation based on approximate composite GPs, with a specific focus on approximating derivative GPs. 
The GP approximation is achieved by generalizing Hilbert space methods to obtain a reduced-rank representation of the composite covariance function through its spectral decomposition. 
Specifically, we derive and analyze the spectral decomposition of derivative covariance functions and further study their properties theoretically. 
Through our approximations, we reduce the steep cubic computational complexity of exact composite GPs (and derivative GPs) to linear with respect to the number of observations.
Our method thus allows analyzing large data with composite and derivative GPs, which remains infeasible with similarly specified exact GPs. 
In our simulation studies, we show that the proposed GP approximations are superior compared to their exact GP counterparts, both in terms of speed and estimation accuracy for latent variable inputs.

We illustrate our models on a real-world case study depicting the maturation process of blood progenitors cells to erythroid cells. 
Our models promise a full reconstruction of the continuous temporal sequence of cells by refining a discrete quantity denoting only a few experimental time points. 
This continuous temporal ordering holds the potential to better explain the underlying biological process as compared to the experimental time. 
Using our models, we thus provide a strong method for estimating cellular ordering by analyzing full-sized single-cell RNA sequencing data.
That said, our models can be easily applied to latent variable estimation problems arising from other fields of study as well.

\subsection{Limitations and future research}
Our composite and derivative GP approximations depend on the partial covariance structure. In this specification, we assume that the two GPs are independent, thus relaxing the off-diagonal functions in the joint covariance structure.
We hypothesize that even more accurate latent variable estimates can be obtained by approximating the joint derivative GPs. However, this would require further generalizing the Hilbert space approximations to non-positive definite (and possibly asymmetric) kernel functions, which we consider out of scope of the present paper.

The latent variable estimation in our models require a set of noisy observations that are assumed to follow a Gaussian distribution centered around the true latent inputs, with a known measurement error (see Eq. \ref{eqn-latent prior}). 
Depending on the application scenario, this measurement error could be unknown. Thus, future studies should investigate this latent prior specification under unknown measurement error, as well as under entirely different distributional assumptions.

In our real-world case study, we present two approaches to estimate the cellular ordering -- using unspliced and spliced gene expression data as well a combination of spliced gene expressions and its derivative RNA velocity. 
An in-depth comparison between these two approaches would require further case studies using various single-cell data, both
real and simulated.
Future research should investigate the advantages (or disadvantages) in choosing one approach over another in estimating the latent cellular ordering.

\section{Acknowledgments}

This work was partially funded by the Deutsche Forschungsgemeinschaft (DFG, German Research Foundation) via the Collaborative Research Center 391 (Spatio-Temporal Statistics for the Transition of Energy and Transport) – 520388526 and DFG Project 500663361. Additionally, M.Z. was supported by the Else Kröner Fresenius Stiftung (ClinBrain). The authors acknowledge the computational resources provided by the German Network for Bioinformatics Infrastructure – de.NBI. The authors further thank the International Max Planck Research School for Intelligent Systems for supporting S.M and M.Z. We are grateful to Luna Fazio for reading the manuscript and for thoughtful feedback that contributed to improving the paper.

\section{Code availability}
The codes related to model development, simulation studies and the case study can be found here: \url{https://github.com/Soham6298/Latent-Composite-HSGPs}.

\newpage

\renewcommand{\thefigure}{S\arabic{figure}}
\renewcommand{\thetable}{S\arabic{table}}
\renewcommand{\df}{f^{(1)}}
\renewcommand{\dy}{y^{(1)}}
\renewcommand{\bx}{\mathbf{x}}
\renewcommand{\by}{\mathbf{y}}
\renewcommand{\bdy}{\mathbf{y}^{(1)}}
\renewcommand{\Bf}{\mathbf{f}}
\renewcommand{\bg}{\mathbf{g}}
\renewcommand{\bdf}{\mathbf{f}^{(1)}}
\renewcommand{\dsigma}{\sigma^{(1)}}
\renewcommand{\dalpha}{\alpha^{(1)}}
\renewcommand{\brho}{\bm{\rho}}
\renewcommand{\bsigma}{\bm{\sigma}}
\renewcommand{\balpha}{\bm{\alpha}}
\renewcommand{\bdsigma}{\bm{\sigma}^{(1)}}
\renewcommand{\bdalpha}{\bm{\alpha}^{(1)}}
\newcommand{\br}{\mathbf{r}}
\newcommand{\bomega}{\bm{\omega}}
\newcommand{\ba}{\mathbf{a}}
\newcommand{\bb}{\mathbf{b}}

\section*{Supplementary materials}

We present the proofs of the theoretical results stated in the main paper in Section A. We then discuss model inference strategies in Section B followed by convergence diagnostics related to our simulation studies in Section C. The methods used to summarize and present the results from our simulation studies are presented in Section D followed by results related to latent variable uncertainty calibration and covariance function hyperparameter estimation in Sections E and F. Finally, in Section G, we present additional results related to our real-world case study.

\subsection*{A: Derivative kernels and spectral representations}

In the following, we show how the spectral representation of stationary covariance kernels can be used to characterize the smoothness of the kernel and to derive explicit expressions for the covariance and spectral densities of derivative Gaussian processes. By using Bochner’s theorem we prove regularity conditions ensuring differentiability of the kernel, and derive general forms for the derivative kernels and their spectral representations. We then apply these results to the squared exponential and Matérn classes and conclude by identifying the conditions under which derivative kernels remain positive semidefinite and isotropic.

Let $k(\bx, \bx'): \mathbb{R}^p \times \mathbb{R}^p \to \mathbb{R}$ be a continuous, positive-definite covariance function, defining a mean-zero Gaussian process
\[
f(\bx) \sim \mathcal{GP}\left(\mathbf{0},\, k(\bx, \bx')\right).
\]

If $k$ is stationary, that is $k(\bx, \bx') = k(\bx - \bx') = k(\br)$ with $\br = \bx - \bx'$, then by Bochner's theorem \citep{gihman_linear_2004} $k$ is continuous and positive-definite if and only if there exists a finite nonnegative (spectral) measure $S$ on $\mathbb{R}^p$ such that
\begin{equation*}
k(\br) =  \frac{1}{(2\pi)^p}  \int_{\mathbb{R}^p} \exp \left(  i \bomega^T \br \right) \, S(d\bomega).
\end{equation*}
If $S$ is absolutely continuous with respect to the Lebesgue measure, write 
$S(d\bomega) = S_k(\bomega)\,d\bomega$, the function $S_k(\bomega) \ge 0$ is the 
spectral density corresponding to the covariance function $k(\br)$, and the 
Fourier pair is \citep{stein_interpolation_1999, gp_rasmussen_williams_2006}
\begin{equation}
\begin{aligned}
k(\br) &=   \frac{1}{(2\pi)^p} \int_{\mathbb{R}^p} \exp \left( i \bomega^T \br \right)  S_k(\bomega)\, d\bomega, \\
S_k(\bomega) &= \int_{\mathbb{R}^p}  \exp \left(  -i \bomega^T \br \right)  k(\br)\, d\br,
\qquad \bomega \in \mathbb{R}^p. \\
\end{aligned}
\end{equation}    
We proceed to define the kernel associated to the derivative of the process $f(\bx)$. Assume that $f(\bx)$ is a mean zero GP with stationary kernel
$k(\cdot, \cdot)$. Consider the multi-indices $\ba, \bb \in \mathbb{N}_0^p$ where $\mathbb{N}_0^p = \{ (n_1, \ldots, n_p) \mid n_j \in \mathbb{N}_0, j = 1,..., p \}$. Denote by $g^{(\ba)}(\bx) = \partial^{\ba} f(\bx)$ and $g^{(\bb)}(\bx) = \partial^{\bb} f(\bx)$ where $\partial^{\ba}$ represents the following 
\begin{equation}
    \partial^{\ba} f(\bx) =  \frac{ \partial^{|\ba|} f(\bx) }{ \partial^{a_1} \bx_1 \ldots \partial^{a_p} \bx_p },
\end{equation}
where $|\ba| = \sum_{j=1}^p a_j$. 

We define the derivative $k^{(\ba,\bb)}(\bx, \bx')$ as 
\begin{equation}
\label{eq:derivative-kernel}
    k^{(\ba,\bb)}(\bx, \bx') = \operatorname{Cov} \left(g^{\ba}(\bx), g^{\bb}(\bx')\right) = \partial_\bx^{\ba} \, \partial_{\bx'}^{\bb} \, k(\bx, \bx').
\end{equation}

We proceed to prove the following lemma.
\begin{lemma}
\label{lemma:derivatives}
Assume the spectral density is such that $S_k\in L^1\left( \mathbb{R}^p \right)$ and that for some integer $m\ge 1$,
\[
\int_{\mathbb{R}^p} \|\bomega\|^m S_k(\bomega)\,d\bomega<\infty.
\]
Then, the kernel $k\in C^m(\mathbb{R}^p)$ and, for every multi-index $\ba$ with $|\ba|\leq m$,
\[
\partial_{\br}^{\ba} k(\br)
= \int_{\mathbb{R}^p} (i\bomega)^{\ba} \exp \left( i \bomega^T \br \right) S_k(\bomega)\,d\bomega,
\qquad \br\in\mathbb{R}^p, 
\]

where $(i\bomega)^{\ba} \coloneq \prod_{j=1}^p (i\omega_j)^{a_j}$.
\end{lemma}

\begin{proof}
Fix $j\in\{1,\dots,p\}$ and let $e_j$ be the $j$th canonical vector. By the definition of the partial derivative,
\[
\partial_{r_j} k(\br) = \frac{\partial k(\br)}{\partial r_j}
=\lim_{h\to 0}\frac{k(\br+h e_j)-k(\br)}{h}.
\]
Using Bochner's representation theorem we have that
\[
\frac{k(\br+h e_j)-k(\br)}{h}
=  \int_{\mathbb{R}^p} \exp \left( i \bomega^T \br \right) \frac{\exp(ih \bomega_j)-1}{h} S_k(\bomega)\,d\bomega.
\]

Define the function $ Q_h(\br,\bomega) = \exp(i \bomega^T \br) \frac{\exp(i h\bomega_j)-1}{h} S_k(\bomega)$. For each fixed $\bomega$, $\frac{\exp(i h\omega_j)-1}{h}\to i \omega_j$ as $h\to 0$, therefore
$Q_h(\br,\bomega)\to i \omega_j \exp(i \bomega^T \br)S_k(\bomega)$ pointwise in $\bomega$.

Moreover, since $\left| \exp(i \theta) -1 \right| = 2 \sin(\theta/2)$ and using the fact that $\left| \sin(\theta) \right| \leq \theta$ we have
\[
\left|\frac{\exp(i h\omega_j)-1)}{h}\right|
=\frac{2}{|h|}|\sin(h\omega_j/2)|
\le |\omega_j| \quad\text{for all }h\neq 0.
\]
This implies that $|Q_h(\br,\bomega)|\le |\omega_j| S_k(\bomega)$, which is integrable by hypothesis. Applying the dominated convergence theorem shows that 
\begin{equation*}
\begin{aligned}
\frac{\partial k(\br)}{\partial r_j} &= \lim_{h \to 0}\frac{k(\br+h e_j)-k(\br)}{h} \\
&= \lim_{h \to 0} \int  Q_h(\br,\bomega) d\bomega \\
&= \int \lim_{h \to 0}  Q_h(\br,\bomega) d\bomega \\
&= \int  (i \omega_j) \exp \left( i \bomega^T \br \right) S_k(\bomega)\,d\bomega \\
&= \int  \frac{\partial }{\partial r_j}  \exp \left( i \bomega^T \br \right) S_k(\bomega)\,d\bomega. \\
\end{aligned}
\end{equation*}
To prove continuity of $\partial_{r_j}k(\br)$, consider a fixed but arbitrary sequence $\{\br_n\}_{n=1}^\infty$ such that $\br_n \to \br$. Clearly, $i \omega_j \exp(i \bomega^T \br_n)S_k(\bomega)\to i \omega_j \exp \left( i \bomega^T \br \right) S_k(\bomega)$ pointwise in $\bomega$ and is dominated by the integrable function $|\omega_j|S_k(\bomega)\in L^1$. Therefore, we can apply the dominated convergence theorem to show that
\begin{equation*}
\begin{aligned}
\lim_{n \to \infty} \partial_{r_j}k(\br_n) &= \lim_{n \to \infty} \int_{\mathbb{R}^p}  (i \omega_j) \exp \left( i \bomega^T \br_n \right) S_k(\bomega)\,d\bomega \\
&= \int_{\mathbb{R}^p} \lim_{n \to \infty}  (i \omega_j) \exp \left( i \bomega^T \br_n \right) S_k(\bomega)\,d\bomega \\
&= \int_{\mathbb{R}^p} (i \omega_j) \exp \left( i \bomega^T \br\right)  S_k(\bomega)\,d\bomega  = \partial_{r_j}k(\br).
\end{aligned}
\end{equation*}
This shows that $\partial_{r_j}k$ is continuous and $k \in C^1(\mathbb{R}^p)$.

Iterating the same argument $|\ba|$ times, each step multiplies the integrand by a factor $\omega_{j_\ell}$ and is dominated by $\|\bomega\|^{|\ba|}S_k(\bomega)\in L^1$. Therefore for all $\ba$ such that $|\ba|\leq m$,
\[
\partial_{\br}^{\ba}k(\br)=\int_{\mathbb{R}^p} (i\bomega)^{\ba} \exp \left(i \bomega^T \br \right) S_k(\bomega)\,d\bomega,
\]
and $\partial_{\br}^{\ba} k$ is continuous. This proves $k\in C^m(\mathbb{R}^p)$ and the stated formula.
\end{proof}

We can now show that the derivative kernel $k^{(\ba,\bb)}(\br)$ exists and how it is defined in terms of $k(\br)$.

\begin{proposition}[Derivative kernel]
\label{prop:derivative-kernel}
Let $k(\bx, \bx')= k(\br)$, where $\br = \bx - \bx'$, be a stationary covariance kernel on $\mathbb{R}^p$ with spectral density $S_k \in L^1 \left( \mathbb{R}^p \right)$ such that for some integer $m\ge 1$,
\[
\int_{\mathbb{R}^p} \|\bomega\|^m S_k(\bomega)\,d\bomega<\infty.
\]
For multi-indices $\ba, \bb \in \mathbb{N}_0^p$ such that  $|\ba|+|\bb|\leq m$, define
\begin{equation*}
k^{(\ba,\bb)}(\br)
=\operatorname{Cov}\left(g^{\ba}(\bx),g^{\bb}(\bx')\right)
=\partial_\bx^{\ba}\partial_{\bx'}^{\bb}k(\bx,\bx').
\end{equation*}
Then
\begin{equation*}
k^{(\ba,\bb)}(\br) =(-1)^{|\bb|} \partial_{\br}^{\ba+\bb}k(\br),    
\end{equation*}
where $|\bb| = \sum_{j=1}^p b_j$.
\end{proposition}
\begin{proof}
By Lemma~\ref{lemma:derivatives}, differentiating first with respect to $\bx'$ gives
\begin{equation*}
    \partial_{\bx'}^{\bb} k(\bx, \bx') = \int_{\mathbb{R}^p} (-i\bomega)^{\bb} \exp \left( i \bomega^T \br \right) S_k(\bomega)\,d\bomega.
\end{equation*}
Since $\br = \bx - \bx'$, each derivative with respect to $x_j'$ introduces a $-1$ through $\partial_{x_j'} = \partial_{r_j}$. Differentiating again with respect to $\bx$ yields
\begin{equation*}
\label{eq:derivative-kernel-kabk}
\partial_{\bx}^{\ba} \partial_{\bx'}^{\bb} k(\bx, \bx') =  \int_{\mathbb{R}^p}(i\bomega)^{\ba} (-i\bomega)^{\bb} \exp \left( i \bomega^T \br \right) S_k(\bomega)\,d\bomega.
\end{equation*}
 Therefore, 
\begin{equation*}
\partial_{\bx}^{\ba} \partial_{\bx'}^{\bb} k(\bx, \bx') =  (-1)^{|\bb|} \partial_{\br}^{\ba+\bb} k(\br).
\end{equation*}
\end{proof}

We now show an explicit formula for the spectral density of $k^{(\ba,\bb)}(\br)$ in terms of the spectral density of $k(\br)$.

\begin{proposition}[Spectral representation of derivative kernels]
\label{prop:spectral-derivative}
Let $k^{(\ba,\bb)}(\bx, \bx') = k^{(\ba,\bb)}(\br)$ be a stationary covariance kernel, then its spectral density $S^{(\ba,\bb)}(\bomega)$ is given by
\[
S^{(\ba,\bb)}(\bomega)=(i\bomega)^{\ba} (-i\bomega)^{\bb} S_k(\bomega).
\]
\end{proposition}

\begin{proof}
From Proposition~\ref{prop:derivative-kernel}, 
\begin{equation*}
    k^{(\ba,\bb)}(\br) = \int_{\mathbb{R}^p}(i\bomega)^{\ba} (-i\bomega)^{\bb} \exp \left( i \bomega^T \br \right) S_k(\bomega)\,d\bomega.
\end{equation*}

By the uniqueness statement in Bochner's theorem, this implies that
\[
S^{(\ba,\bb)}(\bomega)=(i\bomega)^{\ba}(-i\bomega)^{\bb}S_k(\bomega).
\]

\end{proof}

We now show when Lemma~\ref{lemma:derivatives} holds for the SE and Matérn kernels.

\begin{corollary}[Squared Exponential]
\label{cor:derivative-se-kernel}
Let $k(\br)$ represent the SE kernel given by,
\begin{equation}
\label{eq:se_cov_appendix}
k(\br) = \alpha^2 \exp\left(-\frac{ \|\br\|^2}{2\rho^2}\right) ,\qquad \br\in\mathbb{R}^p,
\end{equation}
then Lemma~\ref{lemma:derivatives} holds for every integer $m \geq 0$. In particular $k\in C^\infty(\mathbb{R}^p)$ and, for all multi-indices $\ba,\bb \in \mathbb{N}_0^p$, 
\begin{equation*}
    S^{(\ba,\bb)}(\bomega)=(i\bomega)^{\ba}(-i\bomega)^{\bb} S_k(\bomega).
\end{equation*}
\end{corollary}
\begin{proof}
The spectral density of $k(\br)$ is the Fourier transform of an unnormalized Gaussian distribution, given by
\begin{equation*}
\begin{aligned}
S_k(\bomega)&= \int_{\mathbb{R}^p}  \exp \left(  -i \bomega^T \br \right)  k(\br)\, d\br \\
&= \int_{\mathbb{R}^p}
\exp(-i\bomega^T \br) \alpha^2 \exp \left(-\frac{1}{2\rho^2} \|\br\|^2\right) \, d\br  \\
&=(2\pi)^{p/2} \alpha^2 \rho^p \exp\left(- \frac{1}{2}  \rho^2 \| \bomega\|^2 \right).    
\end{aligned}
\end{equation*}
Being Gaussian, it has finite polynomial moments of all orders \citep{stein2011fourier}, hence
\begin{equation*}
    \int_{\mathbb{R}^p}\|\bomega\|^m S_k(\bomega)\,d\bomega<\infty\quad\text{for every } m\ge 0.
\end{equation*}
Therefore the hypothesis of Lemma~\ref{lemma:derivatives} is satisfied for any $m$, yielding $k\in C^\infty(\mathbb{R}^d)$ and the stated identity.
\end{proof}

\begin{corollary}[Matérn]
\label{cor:derivative-matern-kernel}
Let $k(\br)$ be the Matern kernel on $\mathbb{R}^p$ with variance $\alpha^2$, smoothness $\nu>0$ and scale $\rho>0$. Then Lemma~\ref{lemma:derivatives} holds for every integer $m$ with $0 \leq m<2\nu$. For all multi-indices $\ba,\bb \in \mathbb{N}_0^p$ with $|\ba|+|\bb|<2\nu$,
\[
S^{(\ba,\bb)}(\bomega)=(i\bomega)^{\ba}(-i\bomega)^{\bb}\,S_k(\bomega)
\]
and $\partial_{\br}^{\ba}k,\partial_{\br}^{\bb}k$ exist and are continuous. In particular, $k\in C^{m}(\mathbb{R}^p)$ for any $m<2\nu$.
\end{corollary}

We first establish the following auxiliary lemma whose proof appears in \cite{stein2011fourier}.

\begin{lemma}
\label{lemma:polar}    
Let $f:\mathbb{R}^p\to\mathbb{R}$ be an integrable function that depends only on the Euclidean norm, i.e. $f(\bomega)=f(\|\bomega\|)$. Then
\begin{equation*}
    \int_{\mathbb{R}^p} f\left( \| \bomega \| \right) \, d\bomega  = \vartheta_{p-1} \int_ 0^\infty f(l) \ell^{p-1} d\ell 
\end{equation*}
where $\vartheta_{p-1}$ denotes the surface area of the $(p-1)$–dimensional unit sphere,
\[
\vartheta_{p-1} = \frac{2\pi^{p/2}}{\Gamma(p/2)}.
\]

In particular if $B_\varepsilon(0)$ denotes the ball of radius $\varepsilon >0$ centered at the origin we have
\begin{equation*}
    \int_{B_\varepsilon(0)} f(\bomega) \, d\bomega= \vartheta_{p-1} \int_ 0^\varepsilon f(\ell) \ell^{p-1} d\ell. 
\end{equation*}
\end{lemma}

\begin{proof}

The Matérn class of covariance functions is given by 
\begin{equation*}
    k(\br) = \alpha^2  \frac{2^{1-\nu}}{ \Gamma(\nu)} \left(  \frac{\sqrt{2 \nu}}{ \rho }  \|  \br \|  \right)^\nu K_\nu\left(   \frac{\sqrt{2 \nu}}{ \rho }  \|  \br \| \right),  
\end{equation*}
where $K$ is a modified Bessel function \citep{abramowitz_handbook_1965}. Its spectral density is given by \citep{guttorp_whittle-matern}
\begin{equation*}
    S_k(\bomega) = \alpha^2 \frac{\Gamma\left( \nu + \frac{p}{2} \right) (2\nu)^\nu \rho^{-2\nu}}{\Gamma\left( \nu \right) \pi^{p/2}} \left( \frac{2\nu}{\rho^2}+ \| \bomega \|^2 \right)^{-(\nu+p/2)}.
\end{equation*}
Let $I_m = \int_{\mathbb{R}^p} \|\bomega\|^{m} S_k(\bomega) \, d \bomega$. We split the domain of integration of $I_m$ into the unit ball $B_1(0) = \{ \bomega \mid  \| \bomega \| \leq 1 \}$ and its complement: 
\begin{equation*}
\begin{aligned}
I_m &= \int_{\mathbb{R}^p} \|\bomega\|^{m} S_k(\bomega) \, d \bomega \\
&= \int_{\|\bomega\|\leq 1} \|\bomega\|^{m} S_k(\bomega) \, d \bomega + \int_{\|\bomega\|> 1} \|\bomega\|^{m} S_k(\bomega) \, d \bomega \\
\end{aligned}
\end{equation*}

We first consider the integral over $B_1(0)$. Since $S_k(\bomega)$ is continuous on the compact set $B_1(0)$, it is bounded by the Extreme Value theorem. Let $M_0=\sup_{\|\bomega\|\leq 1} S_k(\bomega)<\infty$. Applying Lemma~\ref{lemma:polar} we obtain
\begin{equation*}
\begin{aligned}
\int_{\|\bomega\|\leq 1} \|\bomega\|^{m} S_k(\bomega) \, d \bomega  &\leq \int_{\|\bomega\|\leq 1} \|\bomega\|^{m} M_0 \, d \bomega \\
&\leq M_0   \vartheta_{p-1}  \int_0^1 \ell^{m+p-1} d\ell  \\
&\leq M_0 \frac{2 \pi^{p/2}}{\Gamma(p/2)} \frac{1}{m+p} < \infty.
\end{aligned}
\end{equation*}
We now consider the integral on the complement of $B_1(0)$. We can rewrite the following term involved in $S_k(\bomega)$ as $ \frac{2\nu}{\rho^2} + \| \bomega \|^2 = \| \bomega \|^2 \left( \frac{2\nu}{\rho^2} \frac{1}{\| \bomega \|^2} + 1\right)$. 

Since $\| \bomega \| > 1$ implies $\frac{2\nu}{\rho^2} \frac{1}{\| \bomega \|^2} + 1 \geq 1$, we have 
\begin{equation*}
    \begin{aligned}
        \left( \frac{2\nu}{\rho^2} + \| \bomega \|^2 \right)^{-(\nu + p/2 )} &=  \left( \| \bomega \|^2 \left( \frac{2\nu}{\rho^2} \frac{1}{\| \bomega \|^2} + 1\right)  \right)^{-(\nu + p/2 )}  \leq \| \bomega \|^{- 2 \nu -p }.
    \end{aligned}
\end{equation*}
This provides the following tail bound when $\| \bomega \| > 1$
\begin{equation*}
    \begin{aligned}
      S_k(\bomega) = C \left( \frac{2\nu}{\rho^2} + \| \bomega \|^2 \right)^{-(\nu + p/2 )} \leq C \| \bomega \|^{- 2 \nu -p },
    \end{aligned}
\end{equation*}
where $C = \alpha^2 \frac{\Gamma\left( \nu + \frac{p}{2} \right) (2\nu)^\nu \rho^{-2\nu}}{\Gamma\left( \nu \right) \pi^{p/2}} $. Combining this bound and applying Lemma~\ref{lemma:polar} gives
\begin{equation*}
    \begin{aligned}
      \int_{\|\bomega\| > 1} \|\bomega\|^{m} S_k(\bomega) \, d \bomega  &\leq  \int_{\|\bomega\| > 1} C \|\bomega\|^{m - 2 \nu -p } \, d \bomega \\
      & \leq  C \vartheta_{p-1}\int_1^\infty \ell^{m-2\nu - 1} \, d\ell  \\
      & \leq C  \frac{2 \pi^{p/2}}{\Gamma(p/2)} \int_1^\infty \ell^{m - 2 \nu - 1} \, d\ell.
    \end{aligned}
\end{equation*}
The last integral converges if and only if  $ m  < 2 \nu$. 

Combining the two regions shows $I_m<\infty$ when $m<2\nu$, which verifies the moment condition in Lemma~\ref{lemma:derivatives}.

\end{proof}

Even though we can construct $k^{(\ba,\bb)}(\br)$, it does not imply that it represents a covariance kernel. The following proposition provides sufficient and necessary conditions for this to occur. 

\begin{proposition}[Conditions for $k^{(\ba,\bb)}$ to be a covariance kernel]
\label{prop:psd-derivative-kernel}
Assume the kernel $k$ is stationary and isotropic with spectral density $S_k(\bomega)$ and that $|\ba|+|\bb|\le m$. Then the derivative function $k^{(\ba,\bb)}$ is a positive semidefinite covariance kernel if and only if the following set of conditions hold:
\begin{enumerate}
    \item[(1)] $\quad a_j+b_j\ \text{is even for every }j=1,\dots,p$. 
    \item[(2)] The multi-indices $\ba,\bb$ are such that $|\ba|+3|\bb|\equiv 0\pmod 4$. 
\end{enumerate}

In particular, $\ba=\bb$ always satisfies these conditions, so $k_{\ba,\ba}$ is a covariance kernel.
\end{proposition}

\begin{proof}

From Proposition~\ref{prop:spectral-derivative} the candidate for spectral density of $k^{(\ba,\bb)}$ is
\[
S^{(\ba,\bb)}(\bomega)=(i\bomega)^{\ba}(-i\bomega)^{\bb}S_k(\bomega).
\]

To show the necessity and sufficiency of the conditions we will make use of the fact that a stationary kernel is positive semidefinite if and only if its spectral density is real and nonnegative almost everywhere. Instead of working directly with $k^{(\ba,\bb)}(\br)$, we will work with $S^{(\ba,\bb)}(\bomega)$.

We will first show the necessity of the condition (1). We can write the multiplier $(i\bomega)^{\ba}(-i\bomega)^{\bb}$ coordinatewise as:
\[
(i\bomega)^{\ba}(-i\bomega)^{\bb}
=\prod_{j=1}^p i^{a_j}(-i)^{b_j}\omega_j^{a_j+b_j}.
\]

Let $R_j \bomega $ be the operator that shifts the sign of the $j$th element of $\bomega$:
$$ R_j \bomega = (\omega_1,...,- \omega_j, ..., \omega_p)'. $$

Assume that there exists an index $j$ such that $a_j + b_j$ is odd, then the multiplier associated to $R_j \bomega$ is such that:
\begin{equation*}
    \begin{aligned}
        (iR_j\bomega)^{\ba}(-iR_j\bomega)^{\bb} &=  \left( \prod_{l \neq j }^p (i \omega_l)^{a_l}(-i \omega_l)^{b_l} \right) (- i \omega_j)^{a_j}( i \omega_j)^{b_j}  \\
        &= (-1)^{a_j + b_j} \left( \prod_{l \neq j }^p (i \omega_l)^{a_l}(-i \omega_l)^{b_l} \right) ( i \omega_j)^{a_j}(  -i \omega_j)^{b_j} \\
        &= - \prod_{ l = 1 }^p (i \omega_l)^{a_l}(-i \omega_l)^{b_l} \\
        &= - ( i \bomega)^{\ba}(-i \bomega)^{\bb}.
    \end{aligned}
\end{equation*}
Since $R_j \bomega$ and $\bomega$ have the same norm, the hypothesis that $k(\br)$ is isotropic implies that 
\begin{equation*}
    S_k(R_j \bomega) = S_k(\bomega).
\end{equation*}
Therefore,
\begin{equation*}
     \begin{aligned}
     S^{(\ba,\bb)}(R_j \bomega) &=  (iR_j\bomega)^{\ba}(-iR_j\bomega)^{\bb} S_{k}(R_j \bomega)
            = - ( i \bomega)^{\ba}(-i \bomega)^{\bb} S_k(\bomega) 
            = -  S^{(\ba,\bb)}(\bomega). \\
    \end{aligned}
\end{equation*}

This implies that $S^{(\ba,\bb)}(\bomega)$ can not be nonnegative almost everywhere. This shows it is necessary that $a_j + b_j$ is even for all $j=1,...,p$.

We proceed to show the sufficiency of condition(1).  Assume that $a_j + b_j$ is even for all $j = 1,...p$. 
We can rewrite the factor as 
\begin{equation*}
    \begin{aligned}
   (i\bomega)^{\ba}(-i\bomega)^{\bb} &= \prod_{j=1}^p i^{a_j}(-i)^{b_j}\omega_j^{a_j+b_j} =  \prod_{j=1}^p (i)^{a_j}(-i)^{b_j} \prod_{j=1}^p|\omega_j|^{a_j+b_j} \\
    &= i^{|a|}(-i)^{|b|}\prod_{j=1}^p|\omega_j|^{a_j+b_j} = C \prod_{j=1}^p|\omega_j|^{a_j+b_j}, \\
    \end{aligned}
\end{equation*}
where $C = (-1)^{|\bb|} i^{|\ba|+|\bb| }$. The multiplier will be complex valued if and only if $i^{|\ba|+|\bb| }$ is also complex valued, but the only possible values it can attain are $\{-1, 1 \}$. This shows that condition (1) is sufficient for $k^{(\ba,\bb)}(\br)$ to be real valued.

We now show that condition (2) is sufficient and necessary for nonnegativity of $k^{(\ba,\bb)}(\br)$. 

We can rewrite $C = (-1)^{|\bb|} i^{|\ba|+|\bb| } = (-1)^{|\bb|} (-1)^{ 1/2(|\ba|+|\bb|) } = (-1)^{|\bb| + 1/2(|\ba|+|\bb|)}$. Therefore $ C = 1$ if and only if $|\bb| + 1/2(|\ba|+|\bb|)$ is even, which is equivalent to
\begin{equation*}
    \begin{aligned}
    |\bb|+1/2(|\ba|+|\bb|) \equiv 0\pmod 2
\quad  \Longleftrightarrow \quad
|\ba|+3|\bb|\equiv 0\pmod 4,  
    \end{aligned}
\end{equation*}
which is condition (2). This shows conditions (1) and (2) are necessary and sufficient.

Finally, if $\ba=\bb$, then $a_j+b_j=2a_j$ is even for all $j$ and $|\ba|+3|\ba|=4|\ba|\equiv 0\pmod 4$, so $k^{(\ba,\ba)}$ is positive semidefinite.

\end{proof}

If $k^{(\ba,\bb)}$ is a covariance kernel, then it will be stationary since multiplication by $(i\bomega)^{\ba}(-i\bomega)^{\bb}$ does not introduce any dependence on $\bx$ or  $\bx'$. The function $k^{(\ba,\bb)}$ might not be isotropic, even if $k(\br)$ is. To see this assume that $k$ is isotropic with $k(\br) = \varphi(\|\br\|^2)$, with $\varphi(\cdot)$ a sufficiently smooth function and let $ p = 1 , a = 1, b = 0$, then 
\begin{equation*}
    \begin{aligned}
        k^{(1,0)}(\bx, \bx') = \frac{ \partial k(\br) }{\partial \bx} = \frac{ d  }{d r} \varphi(\br^2) = 2\br \varphi'( \br^2).
    \end{aligned}
\end{equation*}

Since $k^{(1,0)}$ is odd it can not be isotropic. 

The main text considers $k^{(1,1)}$ with $p=1$. When $k(\br)$ is isotropic, then $k^{(1,1)}$ is isotropic as well, since
\begin{equation}
\label{eq:kernel_derivative_isotropy}
    \begin{aligned}
        k^{(1,1)}(\bx,\bx') = \frac{ \partial^2 \varphi(\br^2) }{\partial \bx \partial \bx'} = \frac{ \partial \varphi(\br^2) }{\partial \br^2} =  - 2 \varphi'(\br^2) - 4 \br^2 \varphi''(\br^2).
    \end{aligned}
\end{equation}

\cite{solin_hilbert_2020} 

The Hilbert space approximation methods proposed by \citet{solin_hilbert_2020} rely on the assumptions that the covariance function $k(\cdot,\cdot)$ is stationary and isotropic, and that its spectral density can be expressed as a radial function of the form $S_k(\|\bomega\|) = \psi(\|\bomega\|^2)$. If $\psi(\|\bomega\|^2)$ admits a polynomial expansion,
\[
\psi(\|\bomega\|^2) = \sum_{i=0}^\infty a_i \left(\|\bomega\|^2\right)^i,
\]
which holds, for instance, when $\psi$ is analytic, then the corresponding spectral density can be written as
\[
S_k(\|\bomega\|) = \sum_{i=0}^\infty a_i \left(\|\bomega\|^2\right)^i.
\]

Since $S_{k^{(1,1)}}(\bomega) =  \omega^2 S_k(\bomega) $, it follows that 
$S_{k^{(1,1)}}(\bomega)$ also admits a polynomial expansion of the same form. This implies that the Hilbert space approximation framework introduced by \citet{solin_hilbert_2020} is equally valid for derivative covariance functions such as $k^{(1,1)}$.


For any isotropic covariance function, the spectral density depends only on the frequency norm $\|\bomega\|$, so that a function $\psi(t) = S_k(\sqrt{t})$ with $t = \|\bomega\|^2$ can always be defined. The additional assumption that $\psi$ admits a polynomial expansion requires $\psi$ to be analytic in a neighborhood of $t=0$, which in turn implies that $S_k(\|\bomega\|)$ is infinitely differentiable around the origin. This condition is satisfied by the squared exponential covariance, whose spectral density is Gaussian and hence analytic \citep{stein2011fourier}.

Isotropy does not hold in general for $p>1$. In particular, if $a =\mathbf{e_j}, b= \mathbf{e_l}$ it can be shown that
\begin{equation*}
    \begin{aligned}
        k^{(\mathbf{e_j},\mathbf{e_l})} (\br^2) = \frac{ \partial^2 k(\bx,\bx') }{\partial x_j \partial x_l'} =  - \frac{ \partial \varphi( \| \br \|^2)
 }{\partial r_j r_l} =  -2 \delta_{jl} \varphi'(\| \br \|^2) - 4 r_j r_l  \varphi''(\| \br \|^2).
    \end{aligned}
\end{equation*}

The term $r_j r_\ell$ introduces directional dependence, implying that $k_{j\ell}$ is not invariant under rotations of $\br$.
Therefore, even if $k$ is isotropic, the mixed derivative kernels $k_{j\ell}$ with $j \neq \ell$ are not isotropic.

We can however, form isotropic kernels from non isotropic ones. An example of this is the following one, which is closely related to the usual Laplacian operator  \citep{evans_partial_2010}:
\begin{equation*}
\begin{aligned}
    \Lambda_{(\bx, \bx')} \, k(\bx, \bx') &= \sum_{j=1}^p k^{(  e_j, e_j  ) } \\
    &=  \sum_{j=1}^p -2  \varphi'(\| \br \|^2) - 4 r_j^2 \varphi''(\| \br \|^2). \\
    &=  -2 p  \varphi'(\| \br \|^2) - 4 \| \br \|^2 \varphi''( \| \br \|^2), \\
\end{aligned}
\end{equation*}
which depends only on $\|\br\|^2$ and is therefore isotropic. 
It is also possible to construct isotropic derivative kernels from linear combinations of powers of the Laplacian operator,
\begin{equation*}
P(\Delta) = \sum_{j=0}^m a_j\,\Delta^j,    
\end{equation*}
where $\Delta$ denotes the Laplacian and $a_j\in\mathbb{R}$ are constants. Such operators are invariant to rotations and therefore preserve isotropy \citep{folland_1995}.

\subsection*{B: Inference}\label{sec-inference}
In our proposed methods, the primary parameters of interest are the latent inputs $\bx$ and GP hyperparameters $\bm{\theta}$ given data (outputs) $\by_f$ and $\by_g$ and observed noisy inputs $\tilde{\bx}$. 
Our set of hyperparameters include multi-output GP length scales $\brho_f$ and $\brho_g$, GP marginal variances $\balpha_f$ and $\balpha_g$ as well as error variances $\bsigma_f$ and $\bsigma_g$ corresponding to the HSGPs $\Bf$ and $\bg$ respectively. 
For the $d^{th}$ output dimension, we use independent model hyperparameter priors
\begin{equation}\label{eqn-hyperprior}
    \bm{\theta}_d \sim p(\bm{\theta_d}) = p(\rho_{f_d}) \, p(\alpha_{f_d}) \, p(\sigma_{f_d}) \, p(\rho_{g_d}) \, p(\alpha_{g_d}) \, p(\sigma_{g_d}).
\end{equation}
The joint probability density factorizes as
\begin{equation}\label{eqn-jt prob model}
    p(\by_f, \by_g, \bx, \bm{\theta} \mid \tilde{\bx}) = p(\bx \mid \tilde{\bx}) \prod_d^D p(\by_{f_d} \mid \bx, \bm{\theta}_d) \, p(\by_{g_d} \mid \bx, \bm{\theta}_d) \, p(\bm{\theta}_d),
\end{equation}
where $p(\by_{f_d} \mid \bx, \bm{\theta}_d)$ and $p(\by_{f_d} \mid \bx, \bm{\theta}_d)$ denotes the GP-based likelihoods for a single output dimension and $p(\bx \mid \tilde{\bx})$ denotes the prior for the latent $\bx$ implied by the measurement model in Eq. 21 of the main manuscript. 

The independent structure in the joint probability density occurs due to our relaxed specification of the composite GPs $\Bf$ and $\bg$. 
The details of prior specifications used in our experiments are further discussed in Section 4 of the main manuscript. Using Bayes' theorem, we obtain the joint posterior over $\bx$ and $\bm{\theta}$ as
\begin{equation}\label{eqn-post x}
    p(\bx, \bm{\theta} \mid \by, \tilde{\bx}) = 
    \frac{p(\by_f, \by_g, \bx, \bm{\theta} \mid \tilde{\bx})}
    {\int\int p(\by_f, \by_g, \bx, \bm{\theta} \mid \tilde{\bx}) \, d\bx \, d\bm{\theta}}.
\end{equation}
For the special case of derivative HSGPs, we replace the marginal and error variances with $\bdalpha$ and $\bdsigma$ respectively corresponding to the derivative process $\bdf$. 
A key difference for the derivative HSGP is that it also shares the same length scale $\brho$ between both $\Bf$ and $\bdf$. 
This occurs due to the fundamental derivative covariance structure for $\bdf$ which acts as an extension of the chosen base covariance structure for $\Bf$. 
Thus, we modify Eq.\ref{eqn-hyperprior} for the hyperparameters of derivative HSGPs as
\begin{equation}\label{eqn-hyperprior-deriv}
    p(\bm{\theta}_d) = p(\rho_d) \, p(\alpha_d) \, p(\sigma_d) \, p(\dalpha_d) \, p(\dsigma_d).
\end{equation}
The rest of the inference procedure remains similar to Eq.\ref{eqn-jt prob model} and Eq.\ref{eqn-post x} by replacing $\by_g$ with $\bdy$ and modifying $\bm{\theta}$ with derivative GP hyperparameters. 
Posterior samples of $\bx$ and $\bm{\theta}$ for all output dimensions are obtained via MCMC sampling, specifically the adaptive Hamiltonian Monte Carlo (HMC) sampler \citep{Neal2011HMC, brooks_handbook_2011, HoffmanM2014NUTS} implemented in the probabilistic programming language Stan \citep{Stan_carpenter_2017, Stan_guide_2024}.

\subsection*{C: Model convergence}
\begin{figure}[!ht]
    \centering
    \includegraphics[width = \linewidth]{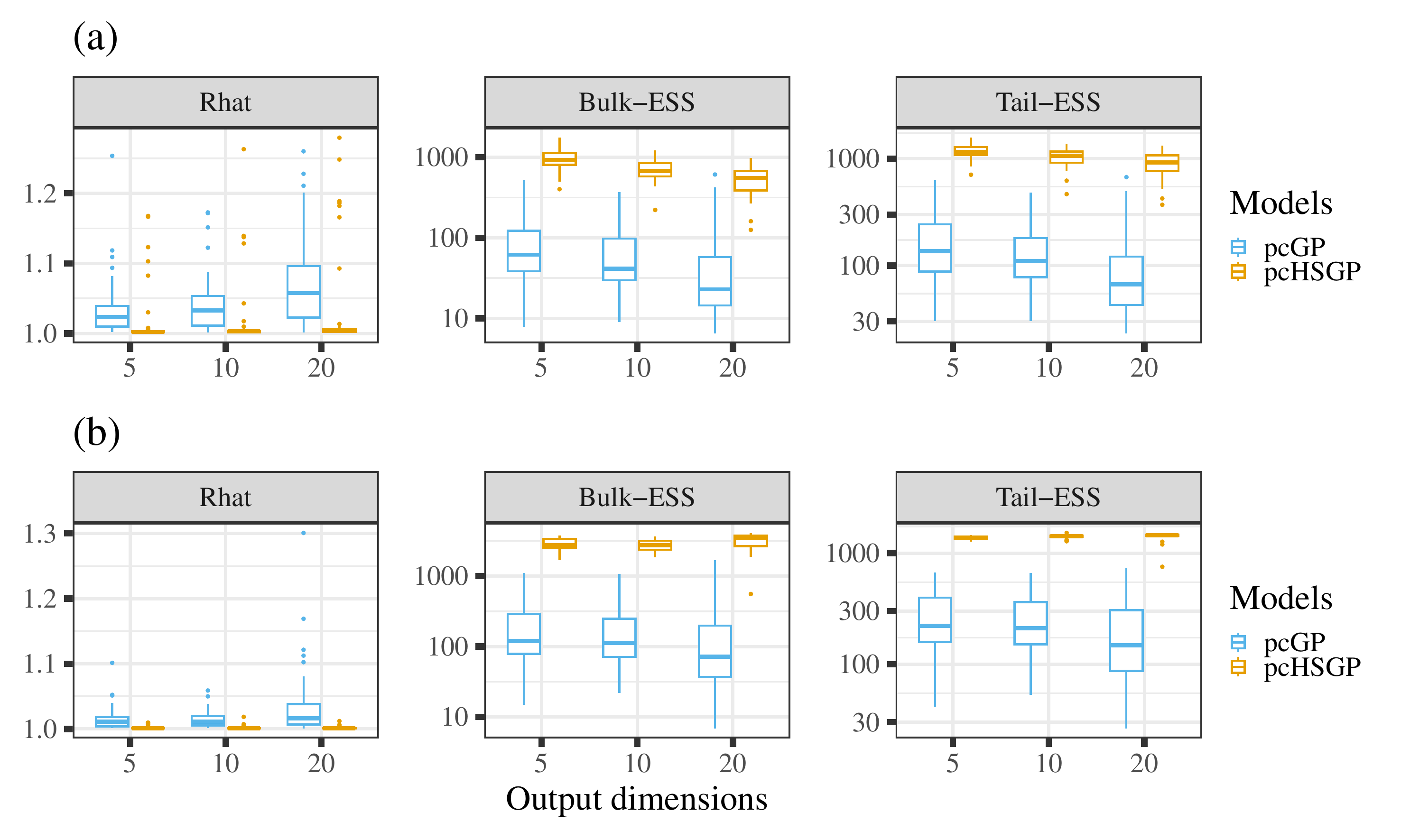}
    \caption{\textit{pcGP data scenario: Convergence check for (a) latent inputs and (b) GP hyperparameters of the exact GPs and HSGPs. The y-axes for Bulk and Tail ESS plots are log10 transformed.}}
    \label{fig:ihsgp-valid}
\end{figure}
We investigate MCMC convergence of our fitted exact GPs and HSGPs for all of the simulation study scenarios discussed before. 
We use standard MCMC sampling diagnostics including latest versions of the scale reduction factor $\widehat{R}$, bulk effective sample size (Bulk-ESS) and tail effective sample size (Tail-ESS)  \citep{RankNorm_Vehtari_etal}. 
A combined check of these measures provide a comprehensive summary of the parameter-specific model convergence. 

In general, $\widehat{R}$ should be very close to 1 and should ideally not exceed 1.01 \citep{RankNorm_Vehtari_etal}. 
Since exact latent GPs have complex posteriors \citep{mukherjee2025hilbert, mukherjee_dgp-lvm_2025}, we additionally consider a more relaxed threshold of 1.1 in our simulation studies. 
Bulk-ESS indicates the reliability of measures of central tendency such as the posterior mean or median. 
Tail-ESS indicates the reliability of the 5\% and 95\% quantile estimates, which are then used to construct credible intervals. 
Both Bulk-ESS and Tail-ESS should have values greater than 100 times the number of MCMC chains (higher is better). 
All of the convergence metrics are considered in combination with comparison to ground truth as another layer of evaluation.
\begin{figure}[!ht]
    \centering
    \includegraphics[width = \linewidth]{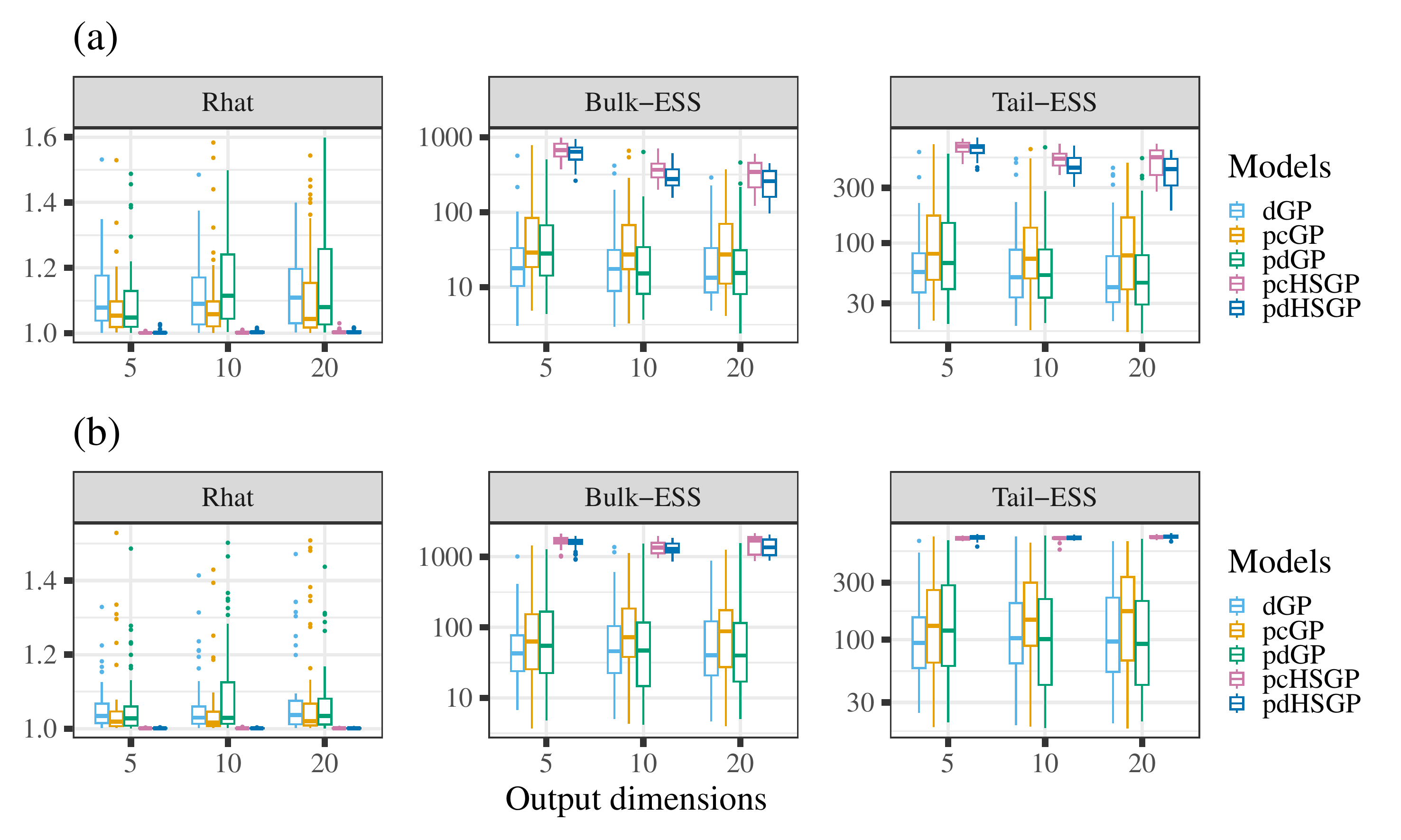}
    \caption{\textit{dGP data scenario: Convergence check for (a) latent inputs and (b) GP hyperparameters of the exact GPs and HSGPs. The y-axes for Bulk and Tail ESS plots are log10 transformed.}}
    \label{fig:deriv-hsgp-valid}
\end{figure}

\begin{figure}[!ht]
    \centering
    \includegraphics[width = \linewidth]{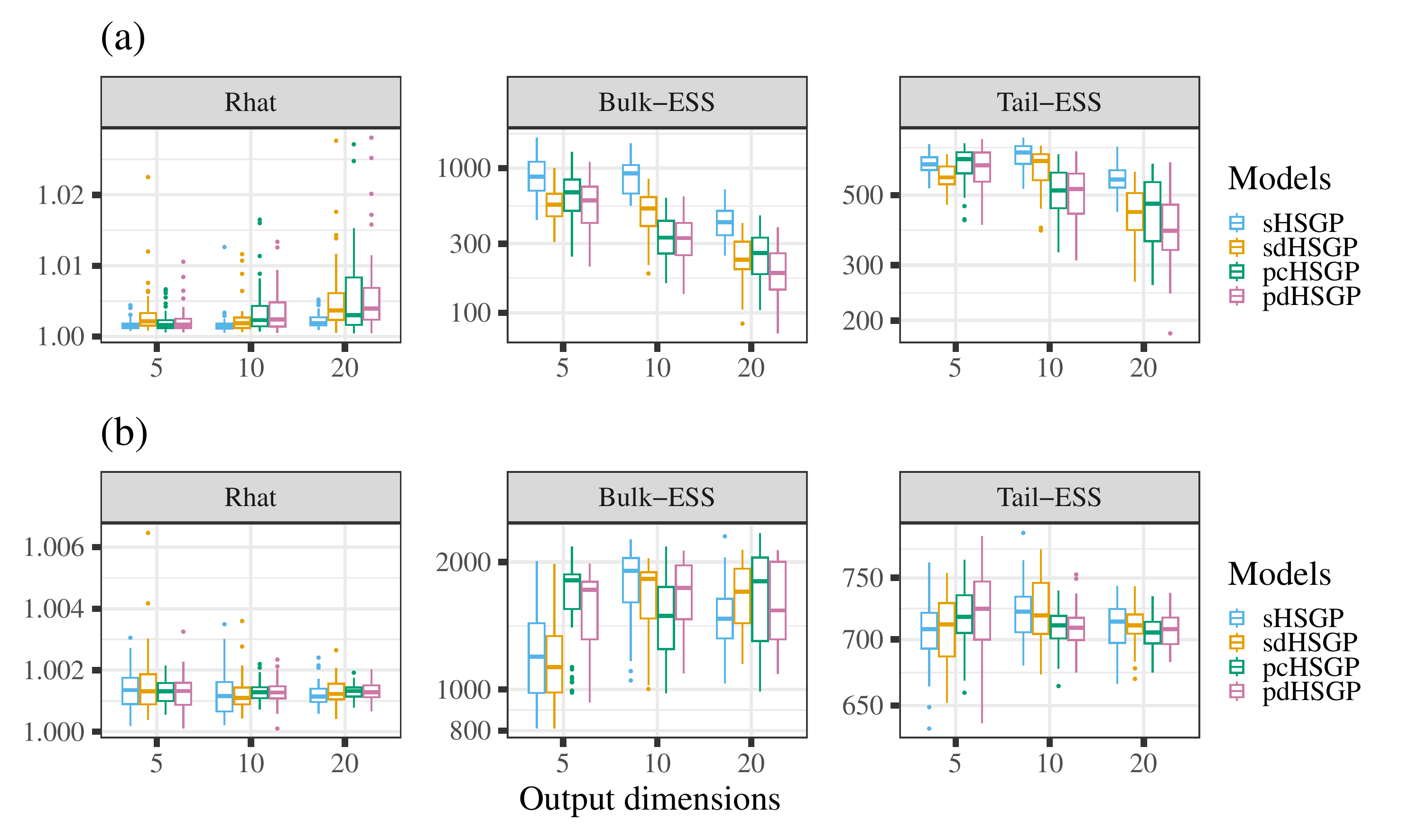}
    \caption{\textit{dGP data scenario: Convergence check for (a) latent inputs and (b) GP hyperparameters of the HSGPs with $N=100$. The y-axes for Bulk and Tail ESS plots are log10 transformed.}}
    \label{fig:deriv-hsgp-n100-valid}
\end{figure}

We show $\widehat{R}$, Bulk-ESS, and Tail-ESS for the latent $\bx$ and GP hyperparameters for the partial composite GP (pcGP) data (in Fig.\ref{fig:ihsgp-valid}) and full derivative GP (dGP) data (in Figures \ref{fig:deriv-hsgp-valid}-\ref{fig:deriv-hsgp-n100-valid}). 
While the exact GPs (including the true data generating processes) reach and exceed the relaxed $\widehat{R}$ threshold of 1.1 for some simulated datasets, the approximate versions pcHSGP and pdHSGP consistently satisfy the much stricter 1.01 threshold of model convergence. 
They subsequently also have much higher Bulk and Tail-ESS as compared to the exact pcGP, pdGP and dGP. 
Overall, based on the diagnostics, the pcHSGPs and pdHSGPs show much more consistent and stable convergence as compared to their exact versions. 
Thus, the pcHSGP and pdHSGPs are suggested both in terms of model convergence diagnostics and practicality in applications to large sample data scenarios.

\subsection*{D: Summary methods}
To evaluate the accuracy of estimating latent variables, we compare posterior samples of the latent input variable $\bx$ denoted by $\bx_{post}$ from each model with their respective ground truth values denoted by $\bx_{true}$. 
Using $\text{RMSE}(\bx_{post}) = \sqrt{\mathbb{E}\left((\bx_{post} - \bx_{true})^2\right)}$, we look at the combined check of bias-variance trade-off in estimating the latent variables. 
We compute the posterior RMSE from all fitted models under each simulation scenarios (two scenarios with sample size $N = 20$ and one with $N = 100$) with output dimensions ($D= 5, \text{ }10\text{ } \text{and}\text{ } 20$). 
We prefer models that provide both low RMSE indicating posterior mean estimates close to the ground truth as well as high precision.

To analyze the results from our experiments, we use a multilevel analysis of variance model (ANOVA) by using brms \citep{burkner_brmS_2017, burkner_brms_advanced_2017}, which disentangles the various components of our simulation study design. 
With $\mu_{resp}$ and $\sigma_{resp}$ being the mean and SD of our response variable, we use
\begin{equation}
    \begin{aligned}
        \bm{\mu}_{resp} &= \mathbf{X}\bm{\beta} + \mathbf{Z}\mathbf{b}+\sum_{\tau}{L}_{\mu_\tau}(t_\tau) &\\
        \bm{\sigma}_{resp} &= \mathbf{X}\bm{\eta} + \mathbf{Z}\mathbf{u}+\sum_{\tau}L_{\sigma_\tau}(t_\tau)
    \end{aligned}
\end{equation}
where $\bm{\beta}, \text{ }\mathbf{b}$ (for $\bm{\mu}_{resp}$) and $\bm{\eta}, \text{ } \mathbf{u}$ (for $\bm{\sigma}_{resp}$) are coefficients at the population and group levels with $\mathbf{X}$ and $\mathbf{Z}$ being the corresponding design matrices. 
The $L_\tau(t_\tau)$ terms denote smooth functions over covariates $t$ fitted via splines. We use these models to summarize the results from our experiments. 

In the simulation studies, the population level design matrix $\mathbf{X}$ contains covariates representing the different models (exact or approximated) and number of output dimensions along with their interaction terms. 
For our pcGP data scenario, we only have two comparing models our simulation scenario: the exact pcGP and approximated pcHSGP. 
Thus we use a two level factor variable accounting for the effect of model choice. In case of the dGP scenario with $N = 20$, we have a five level factor variable denoting our proposed methods pcHSGP, pdHSGP along with their exact versions pcGP, pdGP and the true data generating process dGP. 

In case of the large sample dGP simulation scenario with $N = 100$, we have a four level factor variable denoting the single sHSGP, sdHSGP and the composite pcHSGP and pdHSGP. 
For all of the above cases, we use a three level factor variable depicting the $5$, $10$, and $20$ output dimensions. 
Through $\mathbf{Z}$ we account for the group-level dependency structure in the response induced by fitting multiple models to the same simulated dataset. 
We include a random intercept over datasets as well as corresponding random slopes for the model choices made in the respective simulation scenarios. 

Lastly, to capture the non-linear relation of the response to the ground truth $t$, we introduce thin-plate regression spline \citep{Wood2003splines} terms $L_{\mu_\tau}(t_\tau)$ and $L_{\sigma_\tau}(t_\tau)$. 
The spline terms accounts for any non-linear relationship of the response with respect to the true parameter values $t$. 
To summarize the results of latent variable estimation, we specify posterior RMSE as the response in Section 4.3 (main manuscript). 
In case of model calibration results, we use the log $\gamma$ scores as our response which we discuss in the following section.

\subsection*{E: Simulation based calibration}
Using Simulation-Based Calibration (SBC) \citep{modrakSBCcheck2023, taltsValidaBayesInf2020, fazio2025primedpriorssimulationbasedvalidation}, we test model calibration for estimating latent inputs $\bx$. 
The test is carried out starting with a model, say, $\mathcal{M}_0$. 
Then, we generate $J$ datasets $\by_{(j)}, j = 1,\dots,J$ each of size $N$ from the data generating process that exactly aligns with the model $\mathcal{M}_0$. 
In other words, each individual dataset $\by_{(j)}$ is generated based on a corresponding model parameter draw $\bx_{0_{(j)}}$ from its prior distribution $p(\bx)$. 
We sample from the posterior approximation by fitting $\mathcal{M}_0$ to each of the datasets $\by_{(j)}$ thus resulting in $J$ fitted models $\mathcal{M}^{(j)}$ with respective posteriors $p(\bx \mid \by_{(j)})$ each having $H$ posterior draws $\bx_{(j, h)}$. 
Using $\bx_{0_{(j)}}$ as the ground truth, we then calculate a rank statistic for each univariate posterior quantity $h^*_p(\bx)$ for a specific parameter by counting the number of posterior draws $h^*_p(\bx_{(j, h)})$ that are smaller than $h^*_p(\bx_{0_{(j)}})$. 
The rank statistic $R_{(j)}$ for the model $T_{(j)}$ is then given as 
\begin{equation} \label{eqn-sbc-rank}
    R_{(j)} = \sum_{h=1}^H \mathbb{I}[h^*_p(\bx_{(j, h)}) < h^*_p(\bx_{0_{(j)}})].
\end{equation}
The distribution of these single rank-value per model taken together across all $J$ models is a discrete uniform distribution if the approximate posteriors correspond to the true posteriors. 
Using this property, we assess the correctness of the posterior approximations by testing the rank distribution for uniformity. 
If the rank distribution departs from uniformity, it indicates a mismatch in the data generating process, model implementation, the posterior approximations or a combination of these. 

SBC checks are carried out using a test \cite{modrakSBCcheck2023, sailynojaGraphUniftest2022} based on the probability of $\gamma$ where we observe the most extreme point on the ECDF under the assumption of uniformity. 
In the latter case, the test statistic is given by 
\begin{equation} \label{eqn-gamma-stat}
    \gamma = 2 \quad \min_{j \in \{1, \dots, J+1\}}(\min\{\text{Bin}(R_j \mid J, z_j), 1 - \text{Bin}(R_j-1 \mid J, z_j)\}),
\end{equation}
where $z_j$ is the expected proportion of ranks below $j$ such that $z_j = \frac{j}{J+1}$, $R_j$ is the actual empirical ranks below $j$, $\text{Bin}(R_j \mid J, z_j)$ is the CDF of the Binomial distribution with $J$ trials and the probability of success evaluated at $R$. 
The calculated $\gamma$ scores (presented on the logarithmic scale for ease of visualization) are then compared to a threshold value below which we reject uniformity \citep[the log $\gamma$ score; see][]{sailynojaGraphUniftest2022}. 
The log $\gamma$ scores are advantageous in summarizing large number of parameters, different models as well us various simulation conditions. 
Thus, in this paper, considering the numerous models under various simulation scenarios, we evaluate model calibration using the log $\gamma$ scores.
\begin{figure}[!ht]
   \centering
   \includegraphics[width = \linewidth]{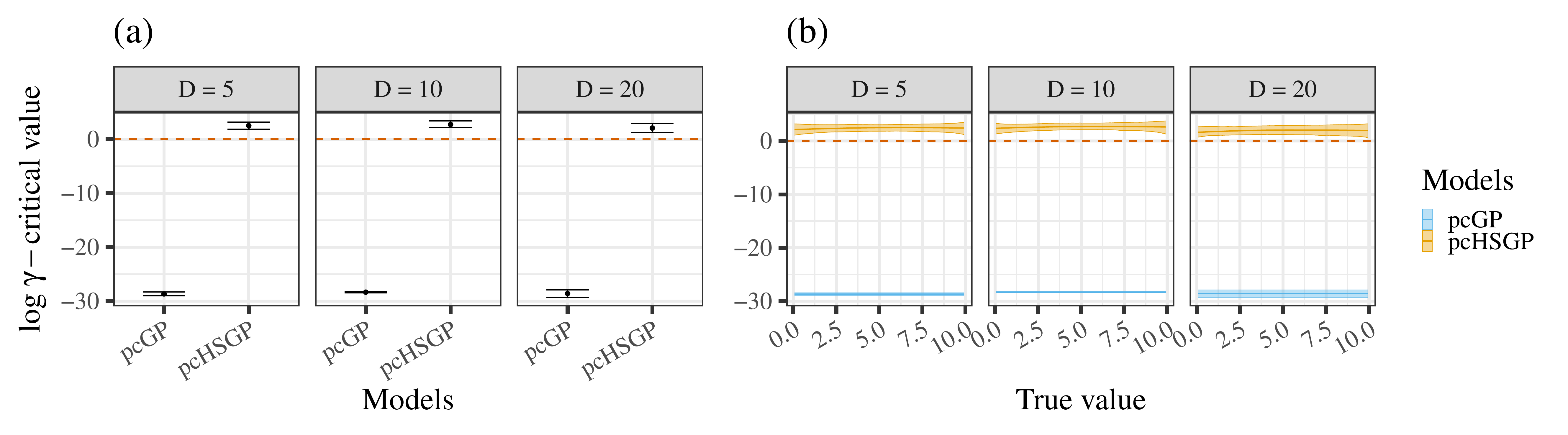}
    \caption{\textit{pcGP data scenario: a) log $\gamma$ scores offset by the 95\% confidence threshold critical value for all the fitted models. The behavior of scores across true latent $x$ values are shown in the right-hand panel. The dashed line denotes the threshold to reject uniformity. b) Shows how the models perform in terms of log $\gamma$ scores with respect to true values across the input space. Figures are shown for output dimensions $D = 5, 10$ and $20$.}}
    \label{fig:pc-data-log-gamma}
\end{figure}

\begin{figure}[!ht]
    \centering
    \includegraphics[width = \linewidth]{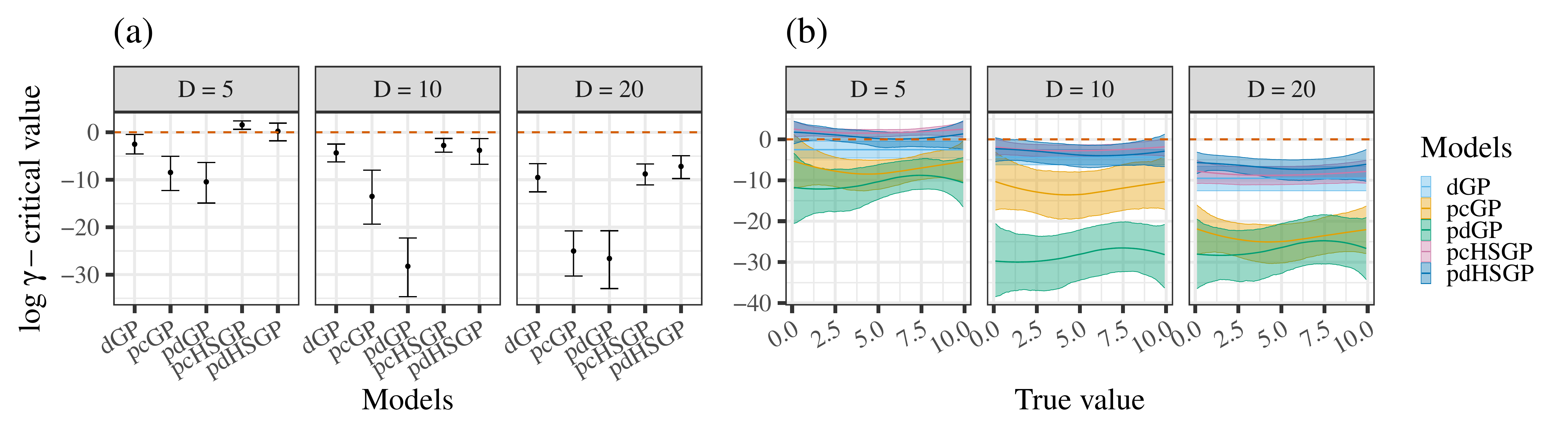}
    \caption{\textit{dGP data scenario: a) log $\gamma$ scores offset by the 95\% confidence threshold critical value for all the fitted models. The behavior of scores across true latent $x$ values are shown in the right-hand panel. The dashed line denotes the threshold to reject uniformity. b) Shows how the models perform in terms of log $\gamma$ scores with respect to true values across the input space. Figures are shown for output dimensions $D = 5, 10$ and $20$.}}
    \label{fig:dgp-data-all-log-gamma}
\end{figure}

To summarize calibration checking results across simulation scenarios, we analyze the log $\gamma$ scores using a multilevel model \citep{burkner_brmS_2017} described in supplementary materials Section C (above). In Figure \ref{fig:pc-data-log-gamma}, we show the predicted log $\gamma$ scores of the exact and approximate models under the pcGP simulation scenario. 
We clearly see that the pcHSGP yields better calibrated results compared to the exact pcGP despite pcGP being the true data generating process.
This result is aligned with the findings in \cite{mukherjee2025hilbert} where the latent exact GPs show various degrees of miscalibration presumably due to the interaction of the sampling procedure with the complex posterior geometries of the models. 
\begin{figure}[!ht]
    \centering
    \includegraphics[width = \linewidth]{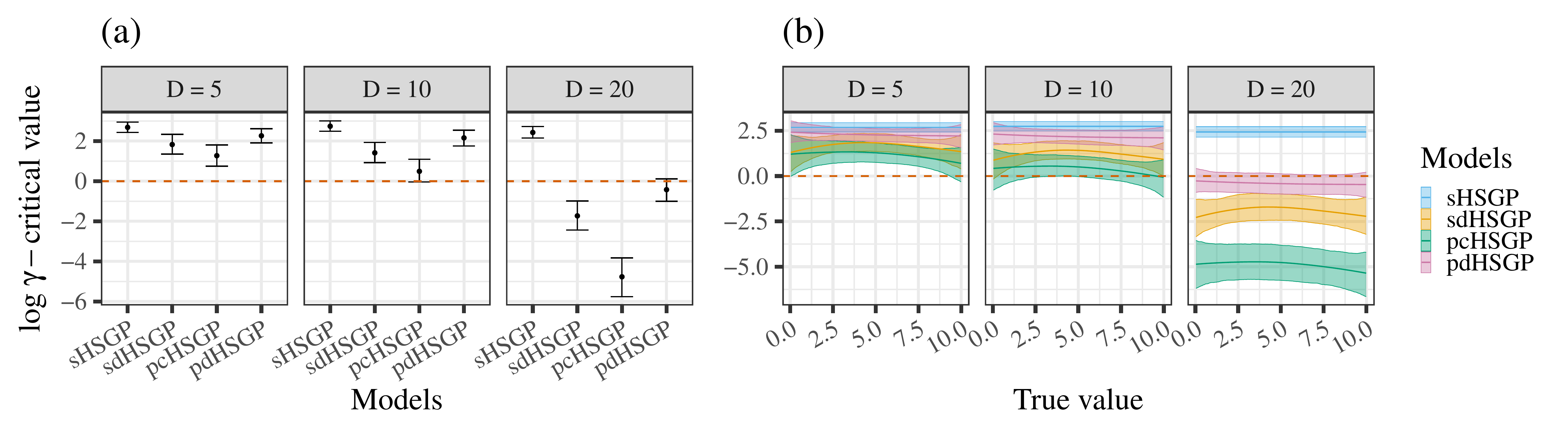}
    \caption{\textit{dGP data scenario: a) log $\gamma$ scores offset by the 95\% confidence threshold critical value for all the fitted models. The behavior of scores across true latent $x$ values are shown in the right-hand panel. The dashed line denotes the threshold to reject uniformity. b) Shows how the models perform in terms of log $\gamma$ scores with respect to true values across the input space. Figures are shown for output dimensions $D = 5, 10$ and $20$.}}
    \label{fig:dgp-data-hsgp-log-gamma}
\end{figure}

In Figure \ref{fig:dgp-data-all-log-gamma}, we detect lack of calibration across all model choices, however, they are so for a few expected reasons. 

For dGP, the reason remains the same as the previous case due to the interaction of complex posterior geometry with the sampling method. The struggle of the samplers with these exact models is also evidenced by the comparatively higher $\widehat{R}$ in the convergence diagnostic results (see Supplementary materials Section C above). For the exact pcGP and pdGP, the miscalibrations additionally are a direct result of relaxing the modeling conditions through using only a partial covariance structure that doesn't fully match the true data generating conditions. 

The extreme miscalibrations for pcGP and pdGP in this case are thus due to a combination of this induced model misspecification as well as complex posterior geometry inherent to latent exact GPs. 
The approximate models pcHSGP and pdHSGP, while not exhibiting the severe miscalibrations of their exact versions, still remain miscalibrated due to the design-induced mis-specifications compared to the true data generating process. These miscalibrations however gets rectified to some extent under higher sample sizes as seen in Figure \ref{fig:dgp-data-hsgp-log-gamma}. For the higher sample size dGP data scenario where we only compare the HSGPs, all models shows good calibrations except for the higher $D = 20$ case. The frequency of data-model mismatch increase with the number fo output dimensions, thus resulting in worse calibration scores for higher $D$. Under these conditions, only the sHSGP and pdHSGP (marginally) passes the posterior calibration checks for the dGP data generating scenario. However, considering the worse (by about 50\%) latent variable recovery (see Section 4.3 Figure 3) compared to pdHSGP, the latter remains the only reasonable choice for these specific data generating conditions. 

\subsection*{F: Hyperparameter recovery}
We show the estimation accuracy for hyperparameters of our proposed methods as well as other comparative models involved in each of the simulation scenarios. 
We, again, present the model evaluation summary using RMSE which combines the model-specific effects on posterior bias and SD for each class of model hyperparameters. We make the comparison based on the class of hyperparameters (for example length-scale) rather than the individual hyperparameters themselves (like $\brho_f$ and $\brho_g$ for composite GPs and $\brho$ for derivative GPs). 
This conscious choice is made since comparability remains valid across all models for our simulation scenarios only through the class of hyperparameters and not the individual hyperparameters.
\begin{figure}[!ht]
    \centering
    \includegraphics[width = \linewidth]{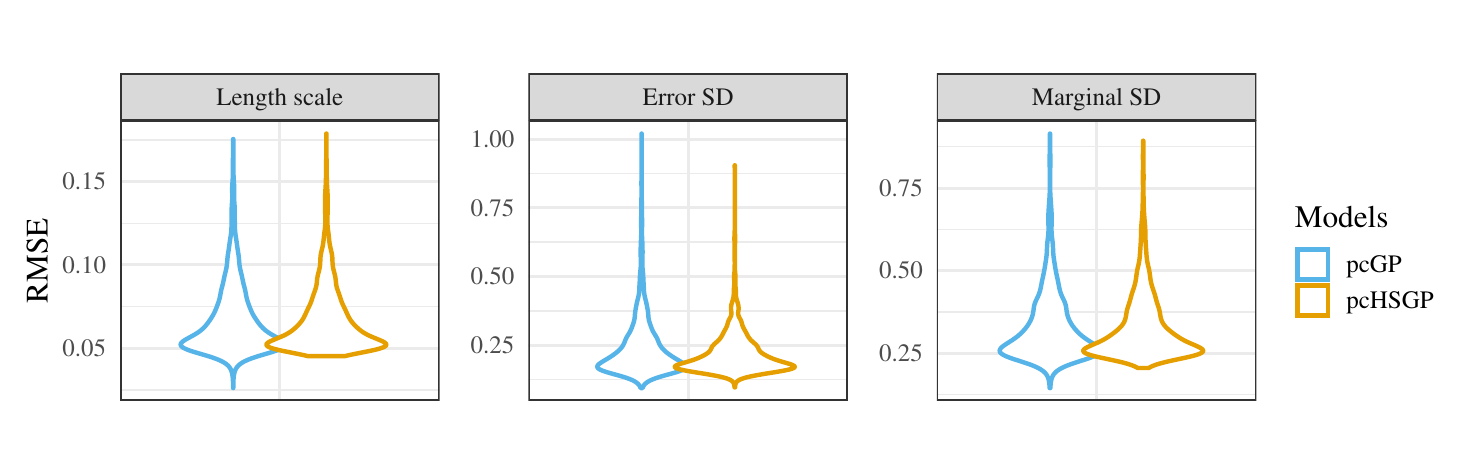}
    \caption{\textit{pcGP data scenario: RMSE on recovery of GP hyperparameters for pcGP and pcHSGP models.}}
    \label{fig:pcgp-data-hyperparams}
\end{figure}

The $D = 20$ case is arguably the most complex case among the choices of output dimensions we have. 
Thus, we decided to showcase only for the $D = 20$ since the results were qualitatively same for all the different choices of output dimensions $D$.
For the pcGP data scenario shown in Figure \ref{fig:pcgp-data-hyperparams}, we don't find any clear differences in hyperparameter estimates between the exact pcGP and the approximate pcHSGP. 
\begin{figure}[!ht]
    \centering
    \includegraphics[width = \linewidth]{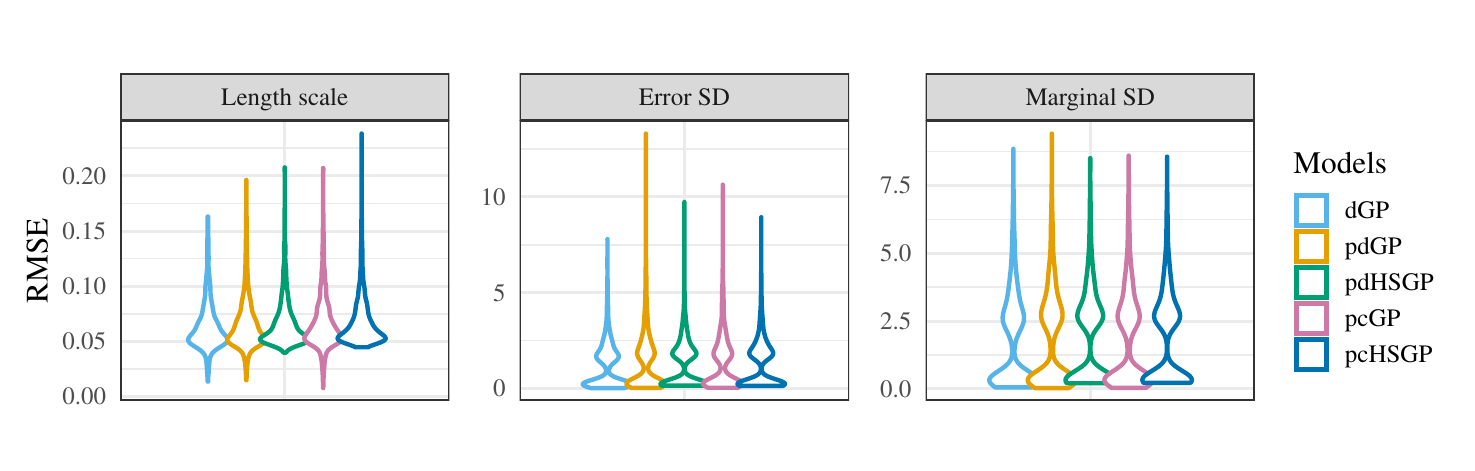}
    \caption{\textit{dGP data scenario: RMSE on recovery of GP hyperparameters for all fitted models.}}
    \label{fig:dgp-data-all-hyperparams}
\end{figure}
In Figure \ref{fig:dgp-data-all-hyperparams}, the approximate pcHSGP and pdHSGP registers similar RMSE results for all of the hyperparameters across all models.
We note that for length scale and error sd, dGP has lesser extreme RMSEs. 
We hypothesize that the partial covariance structure for our proposed HSGPs is likely the primary reason for these phenomena. 

\begin{figure}[!ht]
    \centering
    \includegraphics[width = \linewidth]{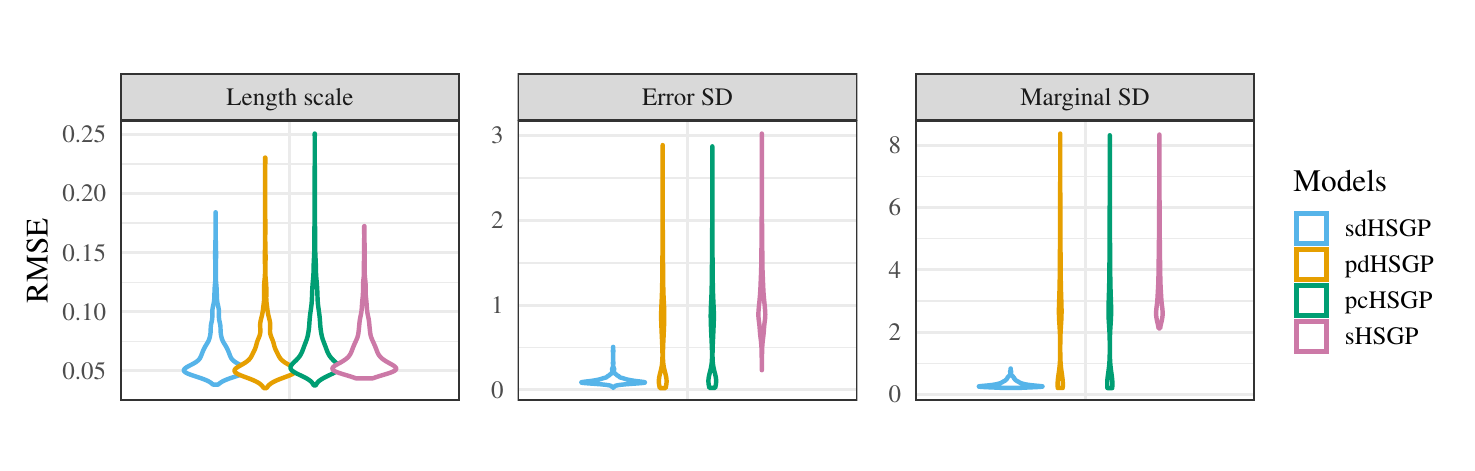}
    \caption{\textit{dGP data scenario ($N = 100$): RMSE on recovery of GP hyperparameters for all fitted models.}}
    \label{fig:dgp-data-hsgps-hyperparams}
\end{figure}

The bimodality in GP marginal SD and error SDs are likely due to the scale differences between GP functions $\Bf, \, \bg$ and the corresponding observations $\by_f, \, \by_g$ (similarly $\bdf$ and $\bdy$ for the derivative cases). 
As the dGP data generating conditions dictate that the derivative $\bdf$ and subsequently $\bdy$ are at a significantly lower scale that $\Bf$ and $\by$, we see this reflected in the RMSEs of marginal and error SDs which modeling this scale imbalance. This behavior is clearly seen in Figure \ref{fig:dgp-data-hsgps-hyperparams} for the marginal and error SDs in sHSGP and sdHSGP which are fitted to $\by$ and $\bdy$ separately. Based on all of the results, the HSGPs and their exact models perform equally when it comes to hyperparameter estimation accuracy for every demonstrated simulation scenario.

\subsection*{G: Additional case study results}
We present additional results to the case study shown in Section 5 (main manuscript) using a smaller set of genes. 
We specifically select 5 genes with highly variable gene expression levels (with a higher degree of non-linearity with respect to experimental time).
\begin{figure}[hbt!]
    \centering
    \includegraphics[width = \linewidth]{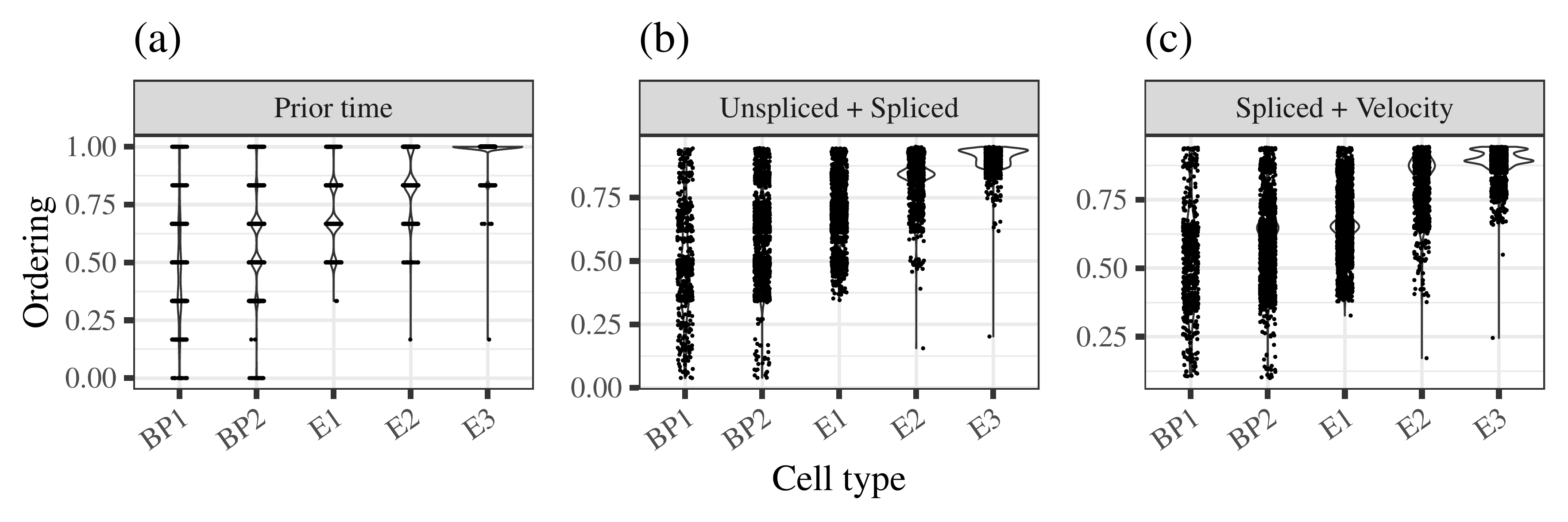}
    \caption{\textit{Case study with 5 genes: a) Distribution of discrete experimental time by cell type. b) Posterior latent continuous cell ordering from modeling unspliced and spliced gene expression using pcHSGP. c) Posterior latent continuous cell ordering from modeling spliced gene expression and RNA velocity using pdHSGP. The x-axis denotes the cell types blood progenitors (BP) and erythroid (E).}}
    \label{fig:case-study-cont-latentx-5gene}
\end{figure}
The list of genes used in the full case study and this shorter study respectively are presented below.
This shorter version of case study with $N = 9815$ and $D = 5$ shows qualitatively similar results as the larger study based on the full $D = 14$ genes suggested in \cite{barile_coordinated_2021}. 
Figures \ref{fig:case-study-cont-latentx-5gene} and \ref{fig:case-study-latentx-scatter-5gene} demonstrate quantitatively similar results as the larger case study. The model fitting times are, however, considerably shorter with pcHSGP and pdHSGP taking 12 and 22 hrs respectively on the same computing resources mentioned in Section 5.
\begin{figure}[hbt!]
    \centering
    \includegraphics[width = \linewidth]{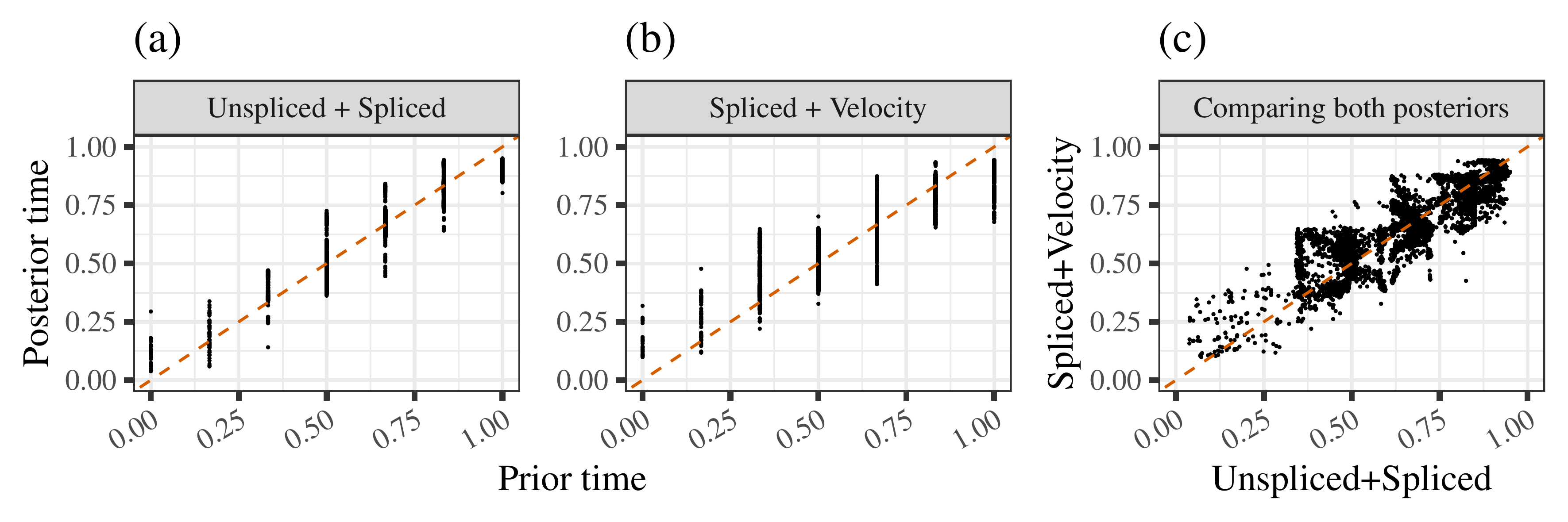}
    \caption{\textit{Case study with 5 genes: Deviations of posterior latent cell ordering from discrete experimental times based on a) modeling unspliced and spliced gene expression using pcHSGP and b) modeling spliced gene expression and RNA velocity using pdHSGP. c) Comparison of posterior latent cell orderings from both the models.}}
    \label{fig:case-study-latentx-scatter-5gene}
\end{figure}

 \begin{table}[hbt!]
     \centering
     \caption{List of gene names that were involved in the cases studies.}
     \begin{threeparttable}
     \begin{tabular}{ccc}
     \toprule
 Gene names & Full case study & Short case study \\
  \midrule
    Blvrb  & \cmark & \cmark \\
    Coro2b & \cmark & \\
    Hba.x  & \cmark & \\
    Hbb.y  & \cmark & \\
    Mllt3  & \cmark & \\
    Myo1b  & \cmark & \\
    Nfkb1  & \cmark & \\
    Phc2 & \cmark & \\
    Rbms2 & \cmark & \cmark \\
    Skap1 & \cmark & \cmark \\
    Smarca2 & \cmark & \\
    Smim1  & \cmark & \cmark \\
    Sulf2 & \cmark & \\
    Yipf5 & \cmark & \cmark \\
  \bottomrule
     \end{tabular}
     \end{threeparttable}
     \label{tab:gene-list}
 \end{table}

\bibliographystyle{abbrvnat}   
\bibliography{refs}

\end{document}